\documentclass{IEEEtran}
\IEEEoverridecommandlockouts
\usepackage{cite}
\usepackage{amsmath,amssymb,amsfonts,amsthm}
\usepackage{algorithmic}
\usepackage{graphicx}
\usepackage{textcomp}
\usepackage{xcolor}
\usepackage{color}
\usepackage{algorithmic}
\usepackage{url}
\usepackage{lipsum}
\usepackage{multirow}
\usepackage[normalem]{ulem}
\usepackage{afterpage}
\usepackage{bbm}
\usepackage{dsfont}
\usepackage[ruled,lined,boxed]{algorithm2e}
\usepackage{mathtools, nccmath}{\tiny {\tiny }}
\usepackage{lipsum}
\usepackage{bm}
\usepackage{threeparttable}
\usepackage{svg}
\usepackage{bbm}
\usepackage{amsmath}
\usepackage{csquotes}
\usepackage[T1]{fontenc}
\usepackage{threeparttable}
\usepackage[top=1.8cm, bottom=1.8cm, left=1.5cm, right=1.5cm]{geometry}
\usepackage{multirow}
\usepackage{makecell}

\DeclareMathAlphabet{\mathbcal}{OMS}{cmsy}{b}{n}
\def\BibTeX{{\rm B\kern-.05em{\sc i\kern-.025em b}\kern-.08em
		T\kern-.1667em\lower.7ex\hbox{E}\kern-.125emX}}
 
\newtheorem{prop}{Proposition}

\newtheorem{theorem}{Theorem}

\newtheoremstyle{iremark}
{\topsep}   
{\topsep}   
{\upshape}  
{0pt}       
{\itshape}  
{.}         
{5pt plus 1pt minus 1pt} 
{\thmname{#1}\thmnumber{ \itshape#2}\thmnote{ (#3)}} 
\theoremstyle{iremark}

\allowdisplaybreaks
\mathchardef\mhyphen="2D
\makeatletter 
\let\myorg@bibitem\bibitem
\def\bibitem#1#2\par{%
	\@ifundefined{bibitem@#1}{%
		\myorg@bibitem{#1}#2\par
	}{%
		\begingroup
		\color{\csname bibitem@#1\endcsname}%
		\myorg@bibitem{#1}#2\par
		\endgroup
	}%
}
\makeatother

\makeatletter
\def\@IEEEtitleabstractindextextfont{%
    \fontsize{10pt}{10pt}\bfseries\selectfont 
}
\makeatother

\begin{document}
{\title{ Exploiting Both Pilots and Data Payloads for \\Integrated Sensing and Communications\\
}
 \author{
 			Chen~Xu,~\IEEEmembership{Graduate Student Member,~IEEE},
 			Xianghao~Yu,~\IEEEmembership{Senior Member,~IEEE}, 
 			Fan~Liu,~\IEEEmembership{Senior Member,~IEEE},\\
 			and Shi Jin, \IEEEmembership{Fellow,~IEEE}
\vspace{-1mm}
	\thanks{
		C. Xu and X. Yu are with the Department of Electrical Engineering, City University of Hong Kong, Hong Kong (e-mail: cxu297-c@my.cityu.edu.hk, alex.yu@cityu.edu.hk).
		
		F. Liu and S. Jin are with the National Mobile Communications Research Laboratory, Southeast University, Nanjing 210096, China (e-mail: f.liu@seu.edu.cn; jinshi@seu.edu.cn).
		
	}
}

\maketitle

\begin{abstract}
	Integrated sensing and communications (ISAC) is one of the key enabling technologies in future sixth-generation (6G) networks.
	 Current ISAC systems predominantly rely on deterministic pilot signals within the signal frame to accomplish sensing tasks. 
	 However, these pilot signals typically occupy only a small portion, e.g., 0.15\% to 25\%, of the time-frequency resources.
	 To enhance the system utility, a promising solution is to repurpose the extensive random data payload signals for sensing tasks.
	 In this paper, we analyze the ISAC performance of a multi-antenna system where both deterministic pilot and random data symbols are employed for sensing tasks. 
	 By capitalizing on random matrix theory (RMT), we first derive a semi-closed-form asymptotic expression of the ergodic linear minimum mean square error (ELMMSE), which evaluates the average sensing error of ISAC systems involving random data payload signals.
	 {\color{black}{Then, we formulate an ISAC precoding optimization problem to minimize the ELMMSE, which is solved via a specifically tailored successive convex approximation (SAC) algorithm.}}
	 To provide system insights, we further derive a closed-form expression for the asymptotic ELMMSE at high signal-to-noise ratios (SNRs). 
	 Our analysis reveals that, compared with conventional sensing implemented by deterministic signals, the sensing performance degradation induced by random signals is critically determined by the ratio of the transmit antenna size to the data symbol length.
	 Based on this result, the ISAC precoding optimization problem at high SNRs is transformed into a convex optimization problem that can be efficiently solved.
	 Simulation results validate the accuracy of the derived asymptotic expressions of ELMMSE and the performance of the proposed precoding schemes. 
	 Particularly, by leveraging data payload signals for sensing tasks, the sensing error is reduced by up to 5.6 dB compared to conventional pilot-based sensing.

\end{abstract}

\begin{IEEEkeywords}
	Convex optimization, ISAC, random matrix theory.
\end{IEEEkeywords}

\section{Introduction}
\bstctlcite{IEEEexample:BSTcontrol}

The sixth-generation (6G) of mobile communication networks are expected to support application scenarios such as autonomous driving \cite{9144301}, low-altitude economy \cite{ye2024integrated,jiang20246G}, and smart cities \cite{liu2023seventy} with its extensive coverage, low latency, high speed, and sensing capabilities, thereby fundamentally reshaping our economy and society. 
Specifically, unlike conventional mobile communication networks, sensing capability is one of the core features of 6G, endowing 6G networks with the ability to perceive the physical world.
Integrated sensing and communications (ISAC), a key enabling technology of 6G and future radar systems, amalgamates the conventional communication and keen sensing functionalities on a unified hardware architecture by reusing time-frequency resources, has attracted significant research attention from both academia and industry~\cite{dong2025CAS}. 

One of the pivotal challenges in ISAC systems is the design of dual-functional waveforms, which integrates communication and sensing signals to concurrently convey useful information to communication users and detect targets presented in physical environments~\cite{liu2024ofdm}. 
Current design philosophy for dual-functional waveforms can be categorized into three types: sensing-centric, communication-centric, and joint designs~\cite{9737357}. 
The communication-centric paradigm aims to enhance the sensing performance within existing communication standards and protocols, while prioritizing communication efficiency~\cite{liu2024ofdm,s22041613,gupta2024affine,wei2024precoding,keskin2023monostatic,rahman2019framework}. 
Due to the compatibility with current wireless communication systems, there have been numerous works investigating sensing performance and precoding design for communication-centric ISAC systems.
For example, the authors of \cite{liu2024ofdm} analyzed the ranging performance of all communication waveforms with cyclic prefix (CP), and proved that the orthogonal frequency division multiplexing (OFDM) is the optimal waveform that achieves the lowest ranging sidelobe.
The authors of \cite{wei2024precoding} investigated the precoding optimization problem according to different design criteria for multiple-input multiple-output (MIMO)-OFDM systems, where a favorable performance tradeoff was achieved by the derived precoding scheme.
To further enhance the sensing and communication performance in MIMO-OFDM ISAC systems, the authors of \cite{gupta2024affine} employed affine precoded superimposed pilot (AP-SIP) signals to improve spectral efficiency and estimation accuracy simultaneously.

Within the current 5G New Radio (NR) standards, base stations (BSs) transmit signals with a specific frame structure to user equipments (UEs)~\cite{wong2017key}. 
Each frame is composed of physical layer reference signals, also referred to as pilot signals, alongside data payload signals~\cite{wong2017key}.
The pilot signal, which is used for conducting channel estimation at the BS side, is typically generated from a deterministic sequence with favorable periodic autocorrelation properties~\cite{3gpp.38.211}.
There have been many works seeking to utilize the pilot signals for various sensing tasks in ISAC systems, such as range and velocity measurement~\cite{10561589,wei20225g}.
However, under current 5G NR standards, pilot signals occupy only 0.15\% to 25\% of the system time-frequency resources~\cite{3gpp.38.211}.
In other words, the data payload signals, which occupy the majority of the system resources, have not yet been fully exploited for sensing tasks. This motivates us to enhance the sensing performance of existing ISAC systems by reusing the remaining at least 75\% data symbols signals in the signal frame.

Nevertheless, different from \emph{deterministic} pilot symbols, data symbols are \emph{randomly} generated according to a certain probability distribution. 
For example, to achieve the channel capacity of an additive white Gaussian noise (AWGN) channel under a power budget constraint, data payload symbols must follow Gaussian distribution~\cite{cover1999elements}.
When random signals are employed for sensing tasks, the challenges are twofold.
First, in conventional wireless communication systems, random transmit signals are normally not directly known at the receiver side. 
Fortunately, prevalent mono-static ISAC BSs inherently have full access to the random data symbols at transceivers, unlocking their potential for random signal-aided sensing.
Second, the randomness of transmit signals transforms classic sensing metrics, such as the Cramer-Rao Bound (CRB) and mean square error (MSE)~\cite{151045}, into random variables. 
This transformation diminishes the effectiveness of these metrics in measuring sensing performance.
One reasonable approach is to employ the expected values of the stochastic versions of CRB or MSE as new sensing performance metrics \cite{xiong2023fundamental,xie2024sensingmutualinformationrandom}.
 
Following this line of research, a performance metric called ergodic linear minimum mean square error (ELMMSE) was defined in \cite{10596930} to evaluate the average MSE of the linear estimator when employing random signals for sensing.
Despite that the ELMMSE is well-defined, it still poses significant challenge of being mathematically intractable. 
More specifically, the precoding design aiming at minimizing the ELMMSE requires solving a difficult constrained stochastic optimization problem with an objective function incorporating random variables.
To cope with this issue, the authors of \cite{10596930} employed a stochastic gradient descent (SGD) algorithm on a mini-batch sample of random data symbols.
Unfortunately, such stochastic optimization algorithms inevitably encounter drawbacks such as computational complexity, hyper-parameter sensitivity, and convergence issues. 
On the other hand, the sensing task in \cite{10596930} is solely implemented by random data payloads. Yet, neglecting the conventional deterministic pilot signals in a practical signal frame structure fails to reveal comprehensive understanding of random signal-assisted sensing and its impact on ISAC precoding design. 
Hence, all these limitations of existing works motivate us to develop tractable performance analysis that leads to efficient precoding design for sensing with both deterministic pilots and random data symbols.

This paper investigates the performance analysis and precoding design for random signal-aided sensing in 6G ISAC systems.
In particular, in addition to the conventional deterministic pilot signals, the random data symbols are reused for sensing tasks.
The main contributions are summarized as follows:
\begin{itemize}
	\item By leveraging the mathematical framework of random matrix theory (RMT), a semi-closed-form asymptotic expression of ELMMSE is derived. Unlike the ELMMSE in \cite{10596930}, the derived asymptotic expression is independent of the random symbols, which significantly simplifies the ISAC precoding design.
	\item Based on the derived semi-closed-form asymptotic expression of ELMMSE, we investigate the precoding design for minimization of the average sensing error while guaranteeing the minimum communication rate under the transmit power budget constraint. Then, a successive convex approximation (SCA)-based algorithm is proposed to obtain a locally optimal solution to the formulated non-convex precoding optimization problem.
	\item To provide deeper system insights, we derive a \emph{closed-form} expression of ELMMSE in the high signal-to-noise ratio (SNR) regimes, unveiling that the degradation in sensing performance incurred by random signals is closely related to {\color{black}{the ratio of the transmit antenna size to the data symbols length. }}Subsequently, the ISAC precoding problem is transformed into a convex optimization problem, for which a globally optimal precoding scheme is achieved.
\end{itemize}

The remainder of this paper is organized as follows: 
Section \ref{sec:SystemModel} introduces the system setting, signal frame structure, and performance metrics. 
The semi-closed-form asymptotic expression of ELMMSE is then derived in Section \ref{sec:AE}.
Section \ref{sec:ISAC_design} discusses the SCA-based algorithm for precoding design, and Section \ref{sec:ISAC_high_SNR}  derives a closed-form asymptotic expression of ELMMSE and the corresponding optimal precoding in high SNR regions.
Section \ref{sec:sim_res} provides the numerical simulations,
and finally Section \ref{sec:conclusion} concludes this paper.

{\emph{Notations}}: 
$\mathbb{R}$ and $\mathbb{C}$ represent the sets of real numbers and complex numbers, respectively.
Matrices are denoted by bold uppercase letters (e.g., $\mathbf{X}$). 
Vectors are represented by bold lowercase letters (e.g., $\mathbf{s}$).
Scalars are denoted by normal font (e.g., $L$). 
$\mathrm{Re}(z)$ and $\mathrm{Im}(z)$ represent the real and imaginary parts of a complex number $z$, respectively. 
The $(i,j)$-th entry of a matrix $\mathbf{Z}$ is denoted as $\left[ \mathbf{Z} \right]_{i,j}$.
$\left(\cdot\right)^T$, $\left(\cdot\right)^*$ and $\left(\cdot\right)^H$ stand for the transpose, conjugate, and Hermitian transpose of a matrix, respectively.
$\mathrm{Tr}(\mathbf{\cdot})$ indicates the trace of a matrix. 
$\mathbf{I}_{N}$ represents an $N \times N$ identity matrix.
The Frobenius norm is written as $\left\| \cdot\right\|_F$.
$\mathbf{A} \succeq \mathbf{0}$ means $\mathbf{A}$ is a positive-semidefinite matrix.
$\cal{O}(\cdot)$ is the big-O notation.
$\mathbb{E}(\cdot)$ represents the expectation of a random variable.
$\mathbb{P}(A)$ denotes the probability of event $A$.
$\mathcal{CN}(\mu,\sigma^2)$ represents the complex Gaussian distribution with mean $\mu$ and variance $\sigma^2$.
$F_n \xrightarrow[]{n \to \infty} F$ indicates that the sequence $F_n$ converges almost surely towards $F$, namely
\begin{equation}
	\begin{aligned}
  		\mathbb{P}\left( \lim_{{n} \rightarrow \infty} F_n=F \right) =1. \nonumber 
  	\end{aligned}
\end{equation}


\section{System Model}\label{sec:SystemModel}

In this section, we first present the considered ISAC signal frame structure, where both deterministic pilot and random data signals are leveraged for sensing tasks. Then, the signal model and performance metrics are introduced.

\subsection{ISAC Signal Frame Structure}
In this paper, we consider a single-user mono-static MIMO ISAC system where the ISAC BS is equipped with $N_{\mathrm{t}}$ transmit and $N_\mathrm{r}$ receive antennas, respectively. 
As shown in Fig. \ref{fig:ISAC}, 
the sensing and communication tasks of the ISAC BS are defined as: 
1) serving a downlink communication UE with $N_\mathrm{c}$ antennas through the communication channel $\mathbf{H}_\mathrm{c}  \in \mathbb{C}^{N_{\mathrm{c}} \times N_{\mathrm{t}}}$;
2) detecting multiple objects whose target impulse response (TIR) matrix is $\mathbf{H}_\mathrm{s} \in \mathbb{C}^{N_{\mathrm{r}} \times N_{\mathrm{t}}} $. {\color{black}{For example, in a typical multiple point targets case, the TIR matrix $\mathbf{H}_\mathrm{s}$ of a uniform linear array (ULA) MIMO system can be written as 
\begin{equation}
	\begin{aligned}
  		\mathbf{H}_\mathrm{s} = \sum_{k=0}^{K-1}\alpha_k \mathbf{b}(\theta) \mathbf{a}^H(\theta),
  	\end{aligned}
\end{equation}
where $K$ is the number of point targets and $\alpha_k \in \mathbb{C}$ represents the $k$-th target's complex reflection coefficient. Moreover, $\mathbf{a}(\theta) \in \mathbb{C}^{N_{\mathrm{t}}\times 1}$ and $\mathbf{b}(\theta) \in \mathbb{C}^{N_{\mathrm{r}}\times 1}$ are steering vector of transmit and receive antenna array, respectively.
}} 
\begin{figure}[t]
	\centering
	\hspace{0cm}\includegraphics[width=0.5\textwidth]{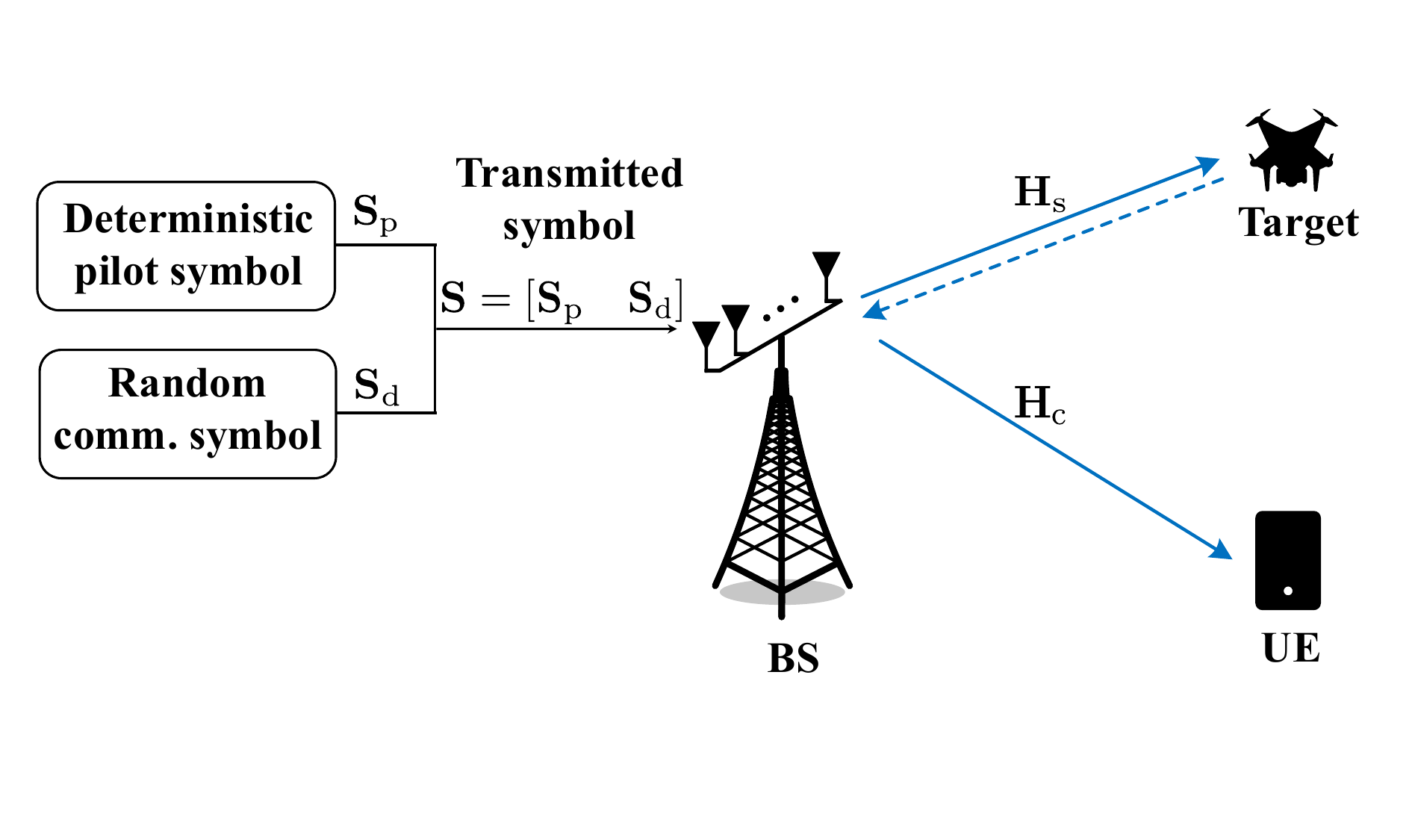}
	\vspace{-11mm}  
	\caption{System model of the considered ISAC system.}
	\vspace{3mm} 
	\label{fig:ISAC}
	
\end{figure}

\begin{figure}[t]
	\centering
	\hspace{0cm}\includegraphics[width=0.48\textwidth]{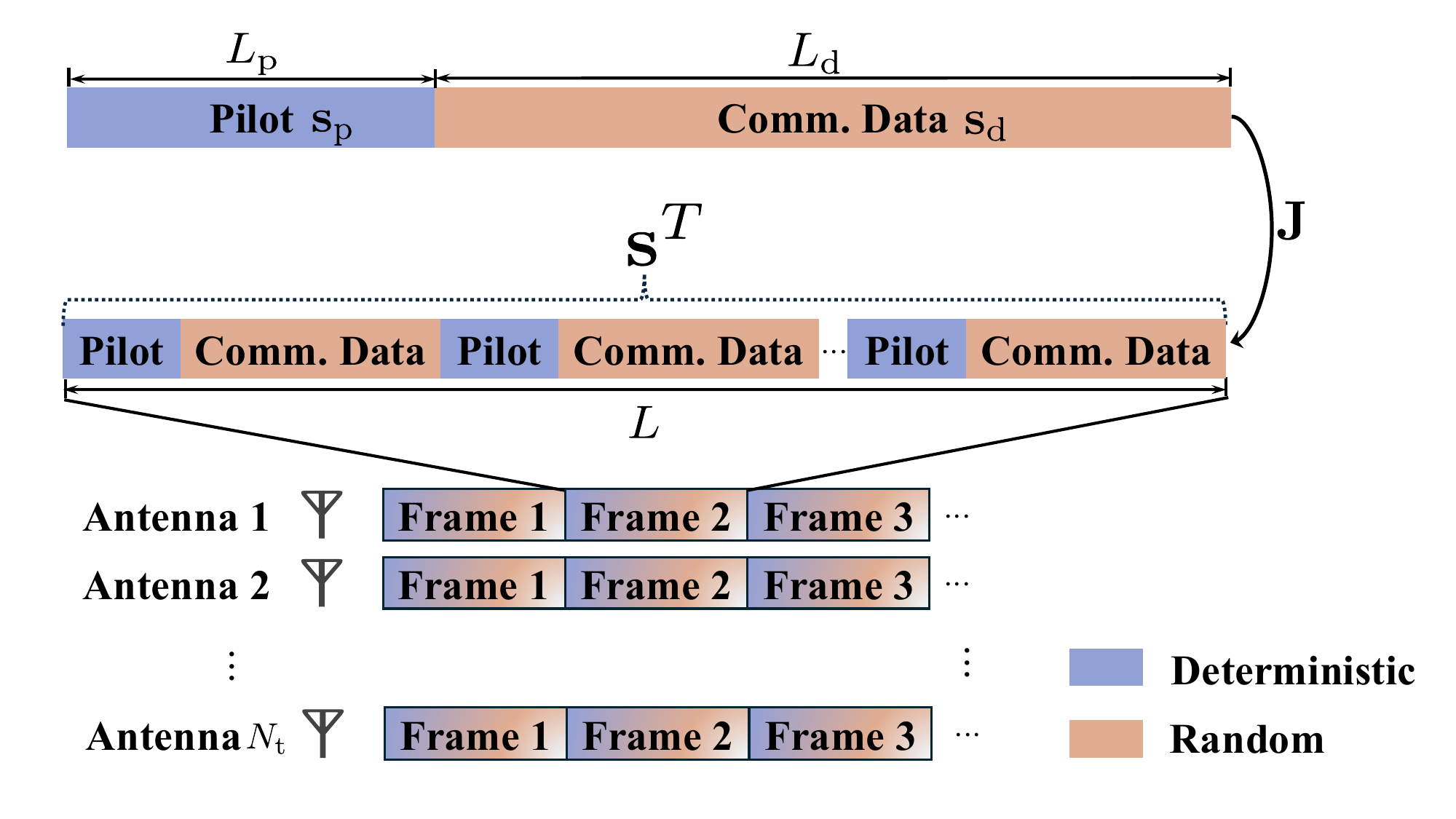}
	\caption{Frame structure, where \enquote{Pilot} and \enquote{Comm. Data} represent the pilot and data symbols, respectively.}
	\label{fig:symbol_stru}
	\vspace{-2mm}  
\end{figure}

In the considered MIMO ISAC system, each transmit antenna emits signals characterized by a consistent frame structure, and the length of each signal frame is denoted as $L$. 
As shown in Fig. \ref{fig:symbol_stru}, the transmit symbols $\mathbf{s} \in  \mathbb{C}^{L \times 1}$ in one frame comprises deterministic pilot symbols ${\mathbf{s}}_\mathrm{p} \in  \mathbb{C}^{L_{\mathrm{p}} \times 1} $ and random data payload symbols ${\mathbf{s}}_\mathrm{d} \in  \mathbb{C}^{L_{\mathrm{d}} \times 1}$. 
In this paper, the communication task is accomplished by random communication symbols. {\color{black}{In contrast to that, both deterministic pilots and random data symbols are leveraged for sensing tasks.}}

In practice, the transmit symbols are interleaved according to predetermined strategies or protocols, such as quasi-periodic placement (QPP) schemes~\cite{1019842}, to minimize the estimation error of the communication channel.
Without loss of generality, we assume $L_\mathrm{p}> N_{\mathrm{t}}$ and $L_\mathrm{d} > N_{\mathrm{t}}$ in this paper.
Therefore, the transmit symbols $\mathbf{s}$ can be represented by means of a permutation matrix $\mathbf{J} \in \mathbb{C}^{L \times L}$~\cite{Petersen2008}, namely
\begin{equation}
	\mathbf{s}^T = \left[
		{\mathbf{s}}_\mathrm{p}^T  \quad {\mathbf{s}}_\mathrm{d}^T \right] \mathbf{J},
\end{equation}
where $\mathbf{J}$ is an orthogonal matrix that has exactly one entry of $1$ in both each row and each column with all other entries being $0$. Fig. \ref{fig:example_J} illustrates an example of permutation matrix $\mathbf{J}$ when $L_\mathrm{p}=2$ and $L_\mathrm{d}=4$.

By concatenating the transmit symbols transmitted by all antennas, we define the deterministic pilot symbol matrix and random data symbol matrix as ${\mathbf{S}}_\mathrm{p} \in \mathbb{C}^{N_{\mathrm{t}} \times L_\mathrm{p}}$ and ${\mathbf{S}}_\mathrm{d} \in \mathbb{C}^{N_{\mathrm{t}} \times L_\mathrm{d}}$, respectively.
Therefore, the MIMO ISAC transmit symbol matrix $\mathbf{S} \in \mathbb{C}^{N_{\mathrm{t}} \times L}$ is given by
\begin{equation}\label{symbol_matrix_stru}
	\mathbf{S} = \left[
		{\mathbf{S}}_\mathrm{p}  \quad {\mathbf{S}}_\mathrm{d} \right] \mathbf{J}.
\end{equation}
To minimize the error of communication channel estimation, the deterministic pilot symbols of different transmit antennas are designed to be orthogonal to each other, namely
\begin{equation}\label{orth_pilot_symb}
	{\mathbf{S}}_\mathrm{p} {\mathbf{S}}_\mathrm{p}^H = \mathbf{I}_{N_{\mathrm{t}}}.
\end{equation}

In contrast, to convey useful information to user, data symbols are drawn from a pre-defined codebook independently, subject to a certain probability distribution. For instance, to achieve the communication capacity of AWGN channels, the data symbols have to follow a complex Gaussian distribution.
Therefore, in this paper we assume that the entries of $\mathbf{S} _\mathrm{d}$ are independent and identically distributed (i.i.d.) complex Gaussian random variables with zero mean and variance $1/{L_\mathrm{d}}$, i.e.,
\begin{equation}
	\begin{aligned}
  		\left[ \mathbf{S} _\mathrm{d} \right]_{i,j} \sim \mathcal{CN} \left( 0,{  \frac{1}{L_\mathrm{d}}} \right),\quad \forall i,j,
  	\end{aligned}
\end{equation}
where the variance ${1}/{L_\mathrm{d}}$ normalizes the energy of data symbols to 1.

\subsection{Signal Model}

\begin{figure}[t]
	\centering
	\hspace{0.3cm}\includegraphics[width=0.5\textwidth]{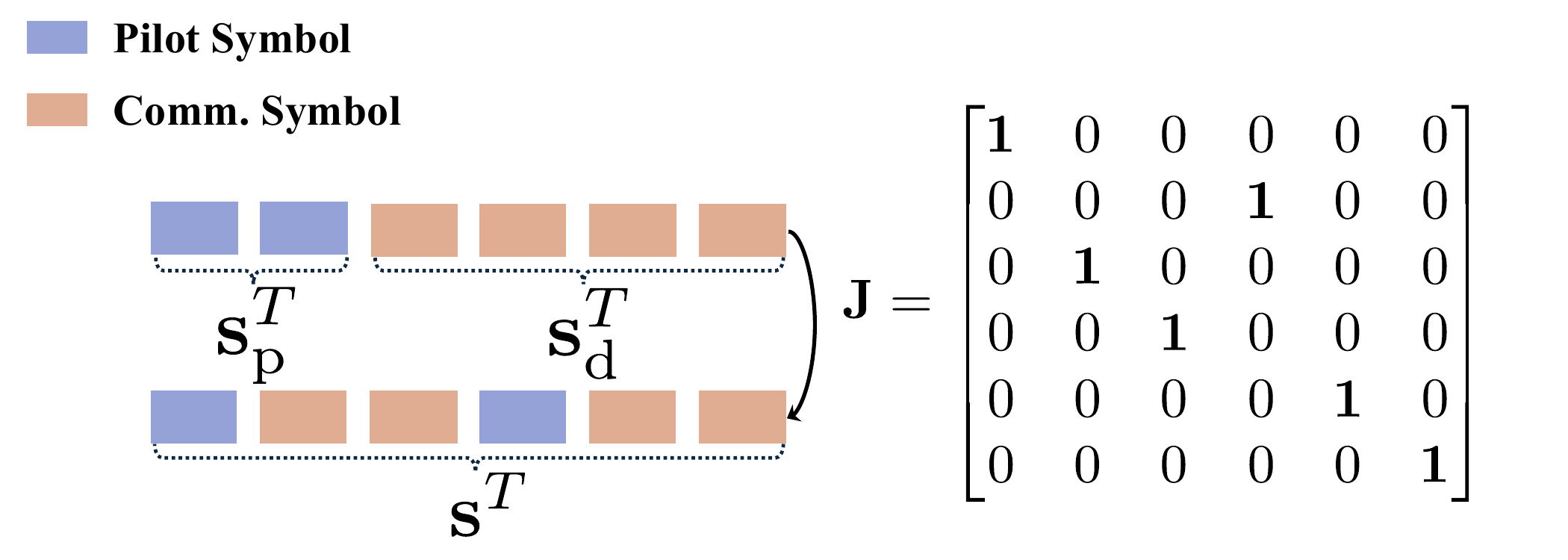}
	\caption{An example of the permutation matrix $\mathbf{J}$ of a frame structure with $L_\mathrm{p}=2$ and $L_\mathrm{d}=4$.}
	\label{fig:example_J}
	\vspace{-2mm}  
\end{figure}

Denote the transmitted ISAC signal matrix as $\mathbf{X} \in \mathbb{C}^{N_{\mathrm{t}} \times L}$, the received signal $\mathbf{Y}_\mathrm{c} \in \mathbb{C}^{N_\mathrm{c} \times L}$ at the communication user is expressed as
\begin{equation}
	\mathbf{Y}_\mathrm{c} = \mathbf{H}_\mathrm{c} \mathbf{X} + \mathbf{Z}_\mathrm{c},
\end{equation}
where the communication channel matrix $\mathbf{H}_\mathrm{c} $ is assumed to be known\footnote{While we consider the pilot symbols in the signal frame, we assume perfect communication CSI in this paper since the focus of this work is to study the utilization of both pilot and data signals on sensing tasks rather than channel estimation. The impact of channel estimation on communication performance, particularly through the use of pilot signals, is deferred to future work for further investigation.}
at the transmitter side of the ISAC BS. 
In addition, $\mathbf{Z}_\mathrm{c} \in \mathbb{C}^{N_\mathrm{c} \times L}$ is the AWGN matrix whose entries are i.i.d. complex Gaussian variables with zero mean and variance $\sigma_\mathrm{c}^2$.

When transmitting useful information to communication users through the communication channel $\mathbf{H}_\mathrm{c}$, the transmit signals $\mathbf{X}$ is meanwhile scattered by the targets in the physical environment and then received at the ISAC BS receiver. 
The received echo signal at the ISAC BS is given by
\begin{equation}
	\mathbf{Y}_\mathrm{s} = \mathbf{H}_\mathrm{s} \mathbf{X} + \mathbf{Z}_\mathrm{s}, \label{signal_model}
\end{equation}
where the correlation matrix of the TIR matrix $\mathbf{R}=\mathbb{E}[\mathbf{H}_\mathrm{s}^H \mathbf{H}_\mathrm{s}]$ is supposed to be known at the ISAC BS. 
Furthermore, $\mathbf{Z}_\mathrm{s} \in \mathbb{C}^{N_{\mathrm{r}} \times L}$ is the AWGN matrix whose entries are i.i.d. complex Gaussian variables with zero mean and variance $\sigma_\mathrm{s}^2$.

Recall the symbol matrix structure in (\ref{symbol_matrix_stru}), and the ISAC signal matrix is specified by
\begin{equation}\label{signal_stru}
	\mathbf{X} = \left[
		{\mathbf{X}}_\mathrm{p}  \quad {\mathbf{X}}_\mathrm{d} \right] \mathbf{J},
\end{equation}
where the pilot signal matrix ${\mathbf{X}}_\mathrm{p}$ and data signal matrix ${\mathbf{X}}_\mathrm{d}$ are given by

\begin{subequations}\label{Xp_Xd}
	\begin{align}
  		{\mathbf{X}}_\mathrm{p} &= \sqrt{\frac{P_{\mathrm{p}}}{N_{\mathrm{t}}}} {\mathbf{S}}_\mathrm{p}, \\
  		{\mathbf{X}}_\mathrm{d} &= \mathbf{W}_\mathrm{} {\mathbf{S}}_\mathrm{d},
  	\end{align}
\end{subequations}
respectively.
$P_{\mathrm{p}}$ is the transmit power of pilot signals.
$\mathbf{W}_\mathrm{} \in \mathbb{C}^{N_{\mathrm{t}} \times N_{\mathrm{t}}}$ represents the precoding matrix for data symbols, and the power budget constraint  regarding $\mathbf{W}$ is given by
\begin{equation}\label{power_budget}
	\begin{aligned}
  		\mathrm{Tr} \left( \mathbf{W}_\mathrm{} \mathbf{W}_\mathrm{}^H\right) \le  P_\mathrm{d},
  	\end{aligned}
\end{equation}
where $P_\mathrm{d}$ is the maximum transmit power of data signals.

\subsection{Performance Metric}

\subsubsection{Sensing Metrics}
In this paper, we employ the LMMSE estimator $\hat{\mathbf{H}}_\mathrm{s}$ to estimate the TIR matrix $\mathbf{H}_\mathrm{s}$, which is given by \cite{Biguesh2006}
\begin{equation}\label{LMMSE_estimator}
	\hat{\mathbf{H}}_\mathrm{s} = \mathbf{Y}_\mathrm{s}(\mathbf{X}^H\mathbf{R}\mathbf{X}+\sigma_\mathrm{s}^2 N_{\mathrm{r}} \mathbf{I}_L)^{-1}\mathbf{X}^H\mathbf{R},
\end{equation}
and the sensing performance is evaluated by estimation MSE $J_{\mathrm{MSE}}$, which is defined as
\begin{equation}\label{MSE_def}
	J_{\mathrm{MSE}} =\mathbb{E} \left[ \Vert\hat{\mathbf{H}}_\mathrm{s} - \mathbf{H}_\mathrm{s} \Vert_F^2 \right].
\end{equation}
By substituting (\ref{LMMSE_estimator}) into (\ref{MSE_def}), we obtain the formulation of the MSE of $\hat{\mathbf{H}}_\mathrm{s}$ as
\begin{equation}\label{LMMSE}
	J_{\mathrm{LMMSE}} = \mathrm{Tr} \left( \mathbf{R}^{-1} + \frac{1}{N_{\mathrm{r}} \sigma_\mathrm{s}^2} \mathbf{X} \mathbf{X}^H \right) ^{-1}.
\end{equation}
Note that $J_{\mathrm{LMMSE}}$ in ($\ref{LMMSE}$) is in principle a random variable due to the randomness introduced by ${\mathbf{X}}_\mathrm{d}$ involved in $\mathbf{X}$, as presented in (\ref{signal_stru}).
The authors of~\cite{10596930} proposed an ergodic estimation error metric, named ELMMSE, which takes the randomness of the transmitted signal into consideration by averaging $J_{\mathrm{LMMSE}}$ over realizations of the random signal $\mathbf{X}_{\mathrm{d}}$. Correspondingly, the ELMMSE is defined as
\begin{equation}\label{ELMMSE}
	J_{\mathrm{ELMMSE}} = \mathbb{E}_{{\mathbf{X}}_\mathrm{d}} \left[ \mathrm{Tr} \left( \mathbf{R}^{-1} + \frac{1}{N_{\mathrm{r}} \sigma_\mathrm{s}^2} \mathbf{X} \mathbf{X}^H \right) ^{-1} \right].
\end{equation}
By substituting (\ref{signal_stru}) and (\ref{Xp_Xd}) into (\ref{ELMMSE}), the ELMMSE can be further recast as
\begin{equation}\label{ELMMSE2}
	\begin{split}
		J_{\mathrm{ELMMSE}}  =  \mathbb{E}_{{\mathbf{S}}_\mathrm{d}} \biggl\{ \mathrm{Tr} \biggl[ \mathbf{R}^{-1} + &\frac{1}{N_{\mathrm{r}} \sigma_\mathrm{s}^2} \biggl(  \frac{P_\mathrm{p}}{N_{\mathrm{t}}}  \mathbf{I}_{N_{\mathrm{t}}} \\
        &  +   \mathbf{W}_\mathrm{} \mathbf{S}_\mathrm{d} \mathbf{S}_\mathrm{d}^H \mathbf{W}_\mathrm{}^H  \biggr) \biggr]^{-1} \biggr\},
	\end{split}
\end{equation}
in which we utilize the properties (\ref{orth_pilot_symb}) and ${\mathbf{J}} {\mathbf{J}}^T =  \mathbf{I}_{L}$.
{\color{black}{It can be observed from (\ref{ELMMSE2}) that the contribution of deterministic pilot signals and random data signals to the sensing performance are determined by matrices $\frac{P_\mathrm{p}}{N_{\mathrm{t}}}  \mathbf{I}_{N_{\mathrm{t}}}$ and $\mathbf{W}_\mathrm{} \mathbf{S}_\mathrm{d} \mathbf{S}_\mathrm{d}^H \mathbf{W}_\mathrm{}^H$, respectively.
}}

\subsubsection{Communication Metric}
{\color{black}{Note that the contributed communication rate is exclusively made by random data signals, as the deterministic pilot signals do not convey any information. Therefore, given a precoding matrix $\mathbf{W}$, the effective communication rate $R$ can be expressed as
\begin{equation}
	\begin{aligned}
  		R_\mathrm{}  = \frac{L_{\mathrm{d}}}{L} C,
  	\end{aligned}
\end{equation}
where $C$ is channel capacity given by \cite{1203154}
\begin{equation}\label{metric_rate}
	C  = \log \det \left(\mathbf{I}_{N_\mathrm{c}}   + \frac{\mathbf{H}_\mathrm{c}  {\mathbf{W}}_\mathrm{} {\mathbf{W}}_\mathrm{}^H \mathbf{H}_\mathrm{c}^H}{\sigma_\mathrm{c}^2}\right).
\end{equation}}}

\section{Asymptotic Expression of ELMMSE}\label{sec:AE}

The ELMMSE defined in (\ref{ELMMSE2}) poses a significant challenge in designing the optimal precoding matrix, as it incorporates the expectation over ${\mathbf{S}}_\mathrm{d}$, thereby transforming the optimization of the precoding matrices $\mathbf{W}$ into a difficult stochastic optimization problem~\cite{10596930}. 
To tackle this challenge, in this section we first establish a lower bound for the ELMMSE.
Subsequently, by leveraging RMT, we derive an asymptotic expression for the ELMMSE. 
This expression shall facilitate the optimization of the precoding matrix in Section \ref{sec:ISAC_design} due to its significantly improved mathematical tractability.

\subsection{Lower Bound of ELMMSE}

To facilitate the evaluation of the ELMMSE, we first derive a lower bound for the ELMMSE of the following proposition.
\begin{prop}
	\label{prop:lowerBound_ELMMSE}
  A lower bound $J_{\mathrm{lb}}$ of ELMMSE defined in (\ref{ELMMSE2}) is given by
  \begin{equation}\label{eq:lb_ELMMSE}
  	J_{\mathrm{ELMMSE}} \ge J_{\mathrm{lb}} = \mathrm{Tr} \left[ \mathbf{R}^{-1} + \frac{1}{N_{\mathrm{r}} \sigma_\mathrm{s}^2} \biggl( \frac{P_\mathrm{p}}{N_{\mathrm{t}}} \mathbf{I}_{N_{\mathrm{t}}} + \mathbf{W}_\mathrm{} \mathbf{W}_\mathrm{}^H \biggr) \right] ^{-1}.
  \end{equation}  
\end{prop}

\begin{IEEEproof}
	Please refer to Appendix \ref{sec:appendix1}.
\end{IEEEproof}

\emph{Remark 1:} We observe from (\ref{eq:lb_ELMMSE}) that the lower bound $J_{\mathrm{lb}}$ is derived by treating $\mathbf{S}_\mathrm{d}$ in (\ref{ELMMSE2}) as a deterministic unitary matrix, namely $\mathbf{S}_\mathrm{d}\mathbf{S}_\mathrm{d}^H= \mathbf{I}_{N_{\mathrm{t}}}$. 
Recall that the entries of $\mathbf{S}_{\mathrm{d}}$ are i.i.d. Gaussian random variables with mean zero and variance $1/L_\mathrm{d}$, and the deterministic unitary hypothesis only holds when $L_\mathrm{d} \to \infty$ according to the Law of Large Numbers.
Consequently, $J_\mathrm{lb}$ tends to be tight exclusively in the large-$L_\mathrm{d}$ regime, which significantly limits its practical applicability. To tackle this issue, in the following we derive an asymptotic ELMMSE expression.

 \subsection{Asymptotic Expression of ELMMSE}\label{sec:lower_bound}
In this paper, we leverage RMT to derive a tighter asymptotic expression of ELMMSE than $J_{\mathrm{lb}}$.
 The primary challenge of dealing with ELMMSE in (\ref{ELMMSE2})  arises from the inclusion of the random matrix ${\mathbf{S}}_\mathrm{d}$. 
 Upon a closer examination of (\ref{ELMMSE2}), we observe that ELMMSE is defined as the trace of a matrix inverse, which is essentially the sum of the eigenvalues of the inverse matrix.
 Therefore, deriving a tractable expression for ELMMSE requires understanding the distribution of eigenvalues of the inverse of a random matrix.
RMT provides a powerful mathematical framework to analyze the eigenvalue distribution of a random matrix and has found its applications in various domains, e.g., quantum physics \cite{Guhr1998} and statistical inference \cite{johnstone2006}.
Hence, we start with several fundamental concepts and theorems in RMT. 

The RMT analyzes the asymptotic behavior of the eigenvalue distribution of matrices whose entries are i.i.d. random variables. 
Given an $N \times N$ Hermitian matrix $\mathbf{A}$, its eigenvalue distribution is described by \emph{empirical spectral distribution} (e.s.d.), which is defined as 
\begin{equation}\label{eq:esd}
	F^{{\bf{A}}}_N (x) = \frac{1}{N}\sum_{i=1}^N \mathds{1}_{\{x|\lambda_i \le x\}}(x),
\end{equation}
 where $x \in \mathbb{R}$ and $\lambda_1,\lambda_2,\dots,\lambda_N$ are eigenvalues of $\mathbf{A}$. 
 Note that the Hermitian property ensures all the eigenvalues of $\mathbf{A}$ are real numbers. 
 $\mathds{1}_A (x)$ denotes the indicator function of set $A$.
For a random matrix, its e.s.d. constitutes a random measure.
The main conclusion drawn from RMT is that, as $N \rightarrow \infty$, some classes of Hermitian random matrices' e.s.d. will converge almost surely to a deterministic distribution named limiting spectral distribution (l.s.d.).


Next, we proceed to derive the asymptotic expression for the ELMMSE by leveraging the framework of RMT. 
By taking a closer look at the definition of ELMMSE in (\ref{ELMMSE2}), we may see that we are essentially interested in the eigenvalue distribution of a Hermitian matrix $\mathbf{G} \in \mathbb{C}^{N_\mathrm{} \times N_\mathrm{}} $ with the following structure:
\begin{equation}
	\mathbf{G} = {\mathbf{A}} + {\mathbf{BSS}}^H {\mathbf{B}}^H,
	\label{eq:matrix_struc}
\end{equation}
where ${\mathbf{A}} \in \mathbb{C}^{N_\mathrm{} \times N_\mathrm{}}$ is a positive semidefinite matrix, ${\mathbf{B}} \in \mathbb{C}^{N_\mathrm{} \times N_\mathrm{}}$ is an arbitrary square matrix, and ${\mathbf{S}} \in \mathbb{C}^{N_\mathrm{} \times M_\mathrm{}} $ is a random matrix whose entries are i.i.d. random variables with zero mean and variance $1/M_\mathrm{}$. 

Notably, it is difficult to calculate the random matrix $\mathbf{G}$'s l.s.d. from its e.s.d. directly. A fundamental tool named as \emph{Stieltjes transform} is used to address this challenge.
The Stieltjes transform encodes the e.s.d. as an analytic function in the complex domain, thereby enabling us to leverage the advantageous properties of analytic functions for studying the e.s.d. itself.
Given an $N \times N$ random matrix $\mathbf{A}$, Stieltjes transform of its e.s.d. is defined as 
\begin{equation}\label{eq:stieltjes_tsfm}
  		m_N^{\mathbf{A}}(z) = \int_{-\infty}^{\infty} \frac{1}{x-z} \mathrm{d} F^{{\bf{A}}}_N (x),
\end{equation}
where $z$ belongs to the complementary to the support of $F^{\mathbf{A}}_N(x)$. One may refer to~\cite{tao2012topics,Couillet_Debbah_2011} for more details on the application of Stieltjes transform in RMT.
It is worth mentioning that $m_N^{\mathbf{A}}(z)$ uniquely determines $F^{\mathbf{A}}_N(z)$ and vice-versa, which forms the foundation for investigating e.s.d. by the Stieltjes transform.

We now present an important theorem, which provides the relationship between the Stieltjes transforms of e.s.d. and l.s.d. of $\mathbf{G}$.

\begin{theorem}\label{theorem:tit}
	For a Hermitian matrix $\mathbf{G}$ defined in (\ref{eq:matrix_struc}), if $\mathbf{B}$ is the square root of a non-negative definite matrix, the Stieltjes transform $m_{N_{\mathrm{}}}^{\mathbf{G}}(z)$ of e.s.d. $F_{N_\mathrm{}}^{\mathbf{G}}(x)$ satisfies
	\begin{equation}\label{tit_convg}
	\begin{aligned}
  		m_{N_{\mathrm{}}}^{\mathbf{G}}(z) - m^{\mathbf{G}}(z) \xrightarrow[]{N,M \to \infty}   0
  	\end{aligned}
	\end{equation}
with $N_{\mathrm{}} / M_{\mathrm{}} \to c\in (0,\infty)$, where $z \in \mathbb{C}\backslash \mathbb{R}^+$. Furthermore, $m^{\mathbf{G}}(z)$ is the Stieltjes transform of $\mathbf{G}$'s l.s.d., which is given by
\begin{equation}\label{eq:tit_mz}
  		m^{\mathbf{G}}(z)= \frac{1}{N_{\mathrm{}}} \mathrm{Tr} \left(  \mathbf{A} +  \frac{\mathbf{B}\mathbf{B}^H} {1+c  e(z)}  - z \mathbf{I}_{N}  \right)^{-1},
\end{equation}
and $e(z)$ is the solution to the following fixed-point equation
\begin{equation}
	\begin{aligned}\label{eq:tit_ez}
  		e(z) = \frac{1}{N_{\mathrm{}}} \mathrm{Tr} \left[ \mathbf{B}\mathbf{B}^H \left(  \mathbf{A} +  \frac{\mathbf{B}\mathbf{B}^H} {1+c  e(z)}  - z \mathbf{I}_{N}  \right)^{-1} \right] .
  	\end{aligned}
\end{equation}
{\color{black} {In particular, $e(z)$ is explicitly given by
\begin{equation}\label{ez_convgence}
	\begin{aligned}
  		e(z)=\lim_{t \to \infty} e_t(z),
  	\end{aligned}
\end{equation}
where $e_0(z)=-{1}/{z}$ and for $t \ge 1$,
\begin{equation}\label{iter_solve_e}
	\begin{aligned}
  		e_t(z)=\frac{1}{N_{\mathrm{}}} \mathrm{Tr} \left[ \mathbf{B}\mathbf{B}^H \left(  \mathbf{A} +  \frac{\mathbf{B}\mathbf{B}^H} {1+c  e_{t-1}(z)}  - z \mathbf{I}_{N}  \right)^{-1} \right].
  	\end{aligned}
\end{equation}

 }}
\begin{proof}
	Please refer to \cite[Theorem~1]{couillet2011deterministic}.
\end{proof}

\end{theorem}

 Theorem \ref{theorem:tit} indicates that the Stieltjes transform of the e.s.d. of $\mathbf{G}$ will converge to a deterministic quantity (the Stieltjes transform of $\mathbf{G}$'s l.s.d.) almost surely when $N_{\mathrm{}}$ and $M$ approach infinity with a fixed ratio $c$.
 With the highly useful Theorem \ref{theorem:tit} at hand, we derive the following proposition giving the asymptotic expression of ELMMSE.

\begin{prop}\label{prop:AE_ELMMSE}
	For ELMMSE defined in (\ref{ELMMSE2}), we have
	\begin{equation}
		\begin{aligned}
  			J_{\mathrm{ELMMSE}} - J_{\mathrm{ae}} \xrightarrow[]{N_{\mathrm{t}},L_{\mathrm{d}}\to \infty}  0.
  		\end{aligned}
	\end{equation}
	The asymptotic expression $J_{\mathrm{ae}}$ is given by
	\begin{equation}\label{eq:Jae}
 	\begin{split}
  		J_{\mathrm{ae}} =		\mathrm{Tr} \biggl[ \mathbf{R}^{-1} + \frac{1}{N_{\mathrm{r}} \sigma_\mathrm{s}^2} \biggl(\frac{P_\mathrm{p}}{N_{\mathrm{t}}}  \mathbf{I}_{N_{\mathrm{t}}}
  					+ \alpha \mathbf{W}_\mathrm{} \mathbf{W}_\mathrm{}^H \biggr) \biggr] ^{-1},
	\end{split}
	\end{equation}
	where
	\begin{equation}\label{alpha_def}
	\begin{aligned}
  		\alpha = \frac{L_{\mathrm{d}}}{L_{\mathrm{d}} + N_{\mathrm{t}} e},
  	\end{aligned}
	\end{equation}
	and $e$ is the constant solution to the following fixed-point equation 
	\begin{equation}\label{e_expression}
    \begin{split}
      e =  \frac{1}{N_{\mathrm{t}} N_{\mathrm{r}}  \sigma_\mathrm{s}^2}  \mathrm{Tr} \biggl\{ \mathbf{W}_\mathrm{} \mathbf{W}_\mathrm{}^H & \biggl[ \mathbf{R}^{-1} + \frac{1}{N_{\mathrm{r}} \sigma_\mathrm{s}^2} \biggl(  \frac{P_\mathrm{p}}{N_{\mathrm{t}}}  \mathbf{I}_{N_{\mathrm{t}}}\\
        &  +   \frac{L_{\mathrm{d}}}{L_{\mathrm{d}} + N_{\mathrm{t}} e} \mathbf{W}_\mathrm{} \mathbf{W}_\mathrm{}^H  \biggr) \biggr]^{-1} \biggl\}.
    	\end{split}
	\end{equation}
{\color{black}{Similar to Theorem \ref{theorem:tit}, $e$ can be explicitly given by
\begin{equation}
	\begin{aligned}
  		e = \lim_{t \to \infty}e_t,
  	\end{aligned}
\end{equation}
where 
\begin{equation}
	\begin{aligned} \label{iter_e_eq}
  		e_t  =  \frac{1}{N_{\mathrm{t}} N_{\mathrm{r}}  \sigma_\mathrm{s}^2}  \mathrm{Tr} \biggl\{ \mathbf{W}_\mathrm{} \mathbf{W}_\mathrm{}^H & \biggl[ \mathbf{R}^{-1} + \frac{1}{N_{\mathrm{r}} \sigma_\mathrm{s}^2} \biggl(  \frac{P_\mathrm{p}}{N_{\mathrm{t}}}  \mathbf{I}_{N_{\mathrm{t}}}\\
        &  +   \frac{L_{\mathrm{d}}}{L_{\mathrm{d}} + N_{\mathrm{t}} e_{t-1}} \mathbf{W}_\mathrm{} \mathbf{W}_\mathrm{}^H  \biggr) \biggr]^{-1} \biggl\}
  	\end{aligned}
\end{equation}
for $t \ge 1$, and the initial value $e_0$ is given by
\begin{equation}\label{e0}
	\begin{aligned}
  		e_0 = \frac{1}{N_{\mathrm{t}} N_{\mathrm{r}}  \sigma_\mathrm{s}^2}  \mathrm{Tr} \biggl[ \mathbf{W}_\mathrm{} \mathbf{W}_\mathrm{}^H & \biggl( \mathbf{R}^{-1} + \frac{1}{N_{\mathrm{r}} \sigma_\mathrm{s}^2}   \frac{P_\mathrm{p}}{N_{\mathrm{t}}}  \mathbf{I}_{N_{\mathrm{t}}} \biggr)^{-1} \biggr].
  	\end{aligned}
\end{equation}}}
	
	\begin{proof}
	Please refer to  Appendix \ref{sec:appendix_AE_ELMMSE}.
	\end{proof}

\end{prop} 
There are several insights involved in Proposition \ref{prop:AE_ELMMSE} that require some further remarks.

\emph{Remark 2:} In comparison to $J_{\mathrm{ELMMSE}}$ defined in (\ref{ELMMSE2}), the most important feature of the derived asymptotic expression $J_{\mathrm{ae}}$ in (\ref{eq:Jae}) is its independence from the statistical properties of the random matrix $\mathbf{S}_{\mathrm{d}}$, which enhances its mathematical tractability. 
Then, the key question to address next is how accurately the asymptotic expression $J_{\mathrm{ae}}$ can approximate $J_{\mathrm{ELMMSE}}$ in practical applications and therefore facilitates efficient system design.
As indicated in Proposition \ref{prop:AE_ELMMSE}, $J_{\mathrm{ae}}$ approaches to ELMMSE with arbitrarily high precision when both $N_\mathrm{t}$ and $L_\mathrm{d}$ approach infinity. 
However, simulation results in Section \ref{sec:sim_res} shall demonstrate that even with smaller values of $N_\mathrm{t}$ and $L_\mathrm{d}$ (e.g., $N_\mathrm{t}=32$ and $L_\mathrm{d}=64$), $J_{\mathrm{ae}}$ approximates ELMMSE with sufficiently high accuracy, thus laying the foundation for our subsequent precoding scheme design in the next section.

\emph{Remark 3:} It can be observed that, by comparing the derived $J_{\mathrm{ae}}$ in (\ref{eq:Jae}) with the lower bound $J_{\mathrm{lb}}$ of the ELMMSE obtained in (\ref{eq:lb_ELMMSE}), an additional factor $\alpha \in (0,1)$ is multiplied by the matrix $\mathbf{W}_\mathrm{} \mathbf{W}_\mathrm{}^H $.
As we mentioned in Remark 1, the lower bound $J_\mathrm{lb}$ is derived by hypothetically treating 
$\mathbf{S}_\mathrm{d}$ as deterministic unitary data symbols, namely $\mathbf{S}_\mathrm{d}\mathbf{S}_\mathrm{d}^H= \mathbf{I}_{N_{\mathrm{t}}}$. 
In this way, $\alpha$ can be interpreted as a factor describing the sensing performance degradation introduced by the randomness of communication symbol matrix $\mathbf{S}_\mathrm{d}$. 
Furthermore, $\alpha$ is readily calculated by a fixed-point iteration in (\ref{iter_e_eq}), making $J_{\mathrm{ae}}$ essentially a semi-closed-form asymptotic expression of ELMMSE, which facilitates efficient performance evaluation.
 {\color{black}{It is also worth noting that, despite the similarity in forms of (\ref{eq:lb_ELMMSE}) and (\ref{eq:Jae}), the underlying analytical frameworks are completely different. 
 In particular, the development of (\ref{eq:lb_ELMMSE}) relies on Jensen's inequality while (\ref{eq:Jae}) is derived by leveraging RMT. This again verifies the motivation and effectiveness of applying RMT for accurate performance analysis of random signal-assisted sensing.}}

\section{ISAC Precoding Design}\label{sec:ISAC_design}

In this section, we investigate optimizing the sensing and communication performance in a joint manner when employing the ISAC signals with the proposed frame structure. 
Particularly, we use the asymptotic expression $J_{\mathrm{ae}}$, rather than $J_{\mathrm{ELMMSE}}$, as our metric for assessing sensing the performance.
The mathematical tractability of $J_{\mathrm{ae}}$ allows us to directly employ a classic yet efficient SCA-based algorithm to optimize the precoding matrix, instead of solving a challenging stochastic optimization problem \cite{10596930}.

\subsection{Problem Formulation}

We aim to minimize the expected sensing error ELMMSE under the constraint of the required communication performance and power budget. The corresponding optimization problem is formulated as
\begin{equation}\label{ISAC_prob}
	\begin{aligned}
     \min_{\mathbf{W}_\mathrm{}} \ &  {J_{\mathrm{ae}}}  \\
     \mathrm{s.t.}  \ &   R_{\mathrm{c}} \ge R_0,\\
     				\ &    \left \lVert \mathbf{W_{\mathrm{}}} \right \rVert_F^2 \le  P_\mathrm{d},
  	\end{aligned} 
\end{equation}
where $R_0$ represents the minimum required communication rate. 
As can be observed, (\ref{ISAC_prob}) is a non-convex optimization problem since the objective function ${J_{\mathrm{ae}}}$ contains the factor $\alpha$, which is determined by the fixed-point equation (\ref{e_expression}) involving the optimization variable $\mathbf{W}$.

\subsection{SCA Algorithm}\label{subsec:SCA}

By comparing problem (\ref{ISAC_prob}) with its counterpart in \cite{10596930}, although the problem is still non-convex, the semi-closed-form expression $J_\mathrm{ae}$ avoids complicated stochastic optimization, enabling the development of low-complexity algorithms.
To address problem (\ref{ISAC_prob}), an SCA technique is proposed to find a locally optimal precoding matrix.
We commence by introducing an auxiliary variable $\mathbf{M}  = \mathbf{W}_\mathrm{}  \mathbf{W}_\mathrm{}^H$ to recast (\ref{ISAC_prob}) as
\begin{equation}\label{ISAC_prob_M}
	\begin{aligned}
     \min_{\mathbf{M}_\mathrm{}} \ & J_{\mathrm{ae}}(\mathbf{M}) \\
     \mathrm{s.t.}  \ &   \frac{L_{\mathrm{d}}}{L} \log \det \left(\mathbf{I}_{N_\mathrm{c}}   + \frac{\mathbf{H}_\mathrm{c}  {\mathbf{M}} \mathbf{H}_\mathrm{c}^H}{\sigma_\mathrm{c}^2}\right) \ge R_0,\\
     				\ &    \mathrm{Tr}(\mathbf{M}) \le P_\mathrm{d}, \\
     				\ & \mathbf{M} \succeq \mathbf{0},
  	\end{aligned} 
\end{equation}
where
\begin{equation}
	\begin{aligned}\label{Jae_M}
  		J_{\mathrm{ae}}(\mathbf{M}) = \mathrm{Tr} \biggl[ \mathbf{R}^{-1} + \frac{1}{N_{\mathrm{r}} \sigma_\mathrm{s}^2} \biggl(\frac{P_\mathrm{p}}{N_{\mathrm{t}}}  \mathbf{I}_{N_{\mathrm{t}}}
  					+ \alpha \mathbf{M} \biggr) \biggr] ^{-1}.
  	\end{aligned}
\end{equation}
{\color{black}{It should be noted that in (\ref{Jae_M}), $\alpha$ is a function of $\mathbf{M}$ rather than a constant, which can only be implicitly defined by (\ref{alpha_def}) and (\ref{e_expression}). Furthermore, it can be verified that this function is non-convex, which consequently makes $J_{\mathrm{ae}}$ a non-convex function of $\mathbf{M}$.}}
Therefore, the SCA technique \cite{razaviyayn2014successive}, which successively solves a sequence of convex approximation problems of the original non-convex problem, can be adopted to obtain a locally optimal solution to the ISAC precoding problem (\ref{ISAC_prob_M}).
Towards that end, we first construct a surrogate function that locally approximates ${J_{\mathrm{ae}}}(\mathbf{M})$ by employing the first-order Taylor expansion $\hat{J}_{\mathrm{ae}}(\mathbf{M}\big|\mathbf{M}_0)$ at point $\mathbf{M}_0$. The Taylor expansion is given by
\begin{equation}
	\begin{split}
  		\hat{J}_{\mathrm{ae}}(\mathbf{M}\big|\mathbf{M}_0) & = J_{\mathrm{ae}}(\mathbf{M}_0) + \\  & \mathrm{Re} \biggl\{ \mathrm{Tr} \biggl[ \left( 2 \frac{\partial J_{\mathrm{ae}}(\mathbf{M})}  
  		{\partial \mathbf{M}^* }  \right)^H   \biggl(\mathbf{M}-\mathbf{M}_0 \biggr) \biggr]   \biggr\}.
  	\end{split}
\end{equation}
To obtain the explicit expression of $\hat{J}_{\mathrm{ae}}(\mathbf{M}\big|\mathbf{M}_0)$, we need to calculate the Wirtinger derivative $\frac{\partial J_{\mathrm{ae}}(\mathbf{M})}  
  		{\partial \mathbf{M}^* }$ \cite{kreutz2009}.
For the sake of simplicity, we first adopt the following notations, 
\begin{subequations}\label{Jae_grad_notation}
	\begin{align}
		&a = \frac{1}{N_{\mathrm{r}} N_{\mathrm{t}} \sigma_\mathrm{s}^2}, \\
		&\beta = \frac{1}{a} - \frac{a \alpha^2  N_{\mathrm{t}}^2}{L_{\mathrm{d}}} \mathrm{Tr}(\mathbf{MTMT}), \\
  		&\mathbf{T} = \biggl[ \mathbf{R}^{-1} + a N_{\mathrm{t}}  \biggl( \frac{P_\mathrm{p}}{N_{\mathrm{t}}}  \mathbf{I}_{N_{\mathrm{t}}}   
  					+ \alpha \mathbf{M} \biggr) \biggr] ^{-1}, \\
  		&\mathbf{G} = \mathbf{T} - a \alpha N_{\mathrm{t}}  \mathbf{TMT}.
  	\end{align}
\end{subequations}
The closed-form expression of Wirtinger derivative $\frac{\partial J_{\mathrm{ae}}(\mathbf{M})} {\partial \mathbf{M}_\mathrm{}^*}$ is given in the following lemma.
\begin{prop} \label{Derivative_J}
The Wirtinger derivative $\frac{\partial J_{\mathrm{ae}}(\mathbf{M})} {\partial \mathbf{M}_\mathrm{}^*}$ is given by
	\begin{equation}
	\begin{aligned}\label{eq:grad_Jae}
  		\frac{\partial J_{\mathrm{ae}}(\mathbf{M})}{\partial \mathbf{M}_\mathrm{}^*}  = \frac{a N_{\mathrm{t}}^2 \alpha^2}{2 \beta L_{\mathrm{d}}} \mathrm{Tr}(\mathbf{T}^2\mathbf{M}) \mathbf{G} - \frac{a N_{\mathrm{t}}\alpha}{2} \mathbf{T}^2.
  	\end{aligned}
\end{equation}
\end{prop}

\begin{IEEEproof}
	Please refer to Appendix \ref{sec:appendix_grad_Jae_proof}.
\end{IEEEproof}

Now we are ready to introduce the SCA algorithm to tackle the ISAC precoding optimization problem (\ref{ISAC_prob_M}).
As shown in \textbf{Algorithm \ref{alg:SAC_ISAC}}, in the $i$-th iteration of the algorithm, the optimization problem is relaxed as
\begin{equation}\label{SCA_subprob}
	\begin{aligned}
     \min_{\mathbf{M}^{(i)}} \ &  \hat{J}_{\mathrm{ae}}(\mathbf{M}^{(i)}\big|\mathbf{M}^{(i-1)})  \\
     \mathrm{s.t.}  \ &   \frac{L_{\mathrm{d}}}{L} \log \det \left(\mathbf{I}_{N_\mathrm{c}}   + \frac{\mathbf{H}_\mathrm{c}  {\mathbf{M}^{(i)}} \mathbf{H}_\mathrm{c}^H}{\sigma_\mathrm{c}^2}\right) \ge R_0,\\
     				\ &    \mathrm{Tr}(\mathbf{M}^{(i)}) \le  P_\mathrm{d}, \\
     				\ & \mathbf{M}^{(i)} \succeq \mathbf{0}.
  	\end{aligned} 
\end{equation}
Note that the subproblem (\ref{SCA_subprob}) in the $i$-th iteration of the SCA algorithm is a convex optimization problem which can be solved by a numerical toolbox such as CVX~\cite{gb08}.
The algorithm terminates when either the predefined precision threshold $\epsilon_{\mathrm{max}}$ or the maximum number of iterations $i_{\mathrm{max}}$ is reached.
Once a locally optimal solution $\mathbf{M}$ is obtained, by denoting its eigenvalue decomposition as $\mathbf{M}=\mathbf{Q}\mathbf{\Sigma}\mathbf{Q}^H$, we can obtain the corresponding locally optimal precoding matrix $\mathbf{W}$ as
\begin{equation}\label{eq:decom_M}
	\begin{aligned}
  		\mathbf{W} = \mathbf{Q}\mathbf{\Sigma}^{\frac{1}{2}}.
  	\end{aligned}
\end{equation}

\begin{algorithm}[t]\label{alg1}
	\caption{SCA-Based Algorithm for Problem (\ref{ISAC_prob_M})}\label{alg:SAC_ISAC}
	\begin{algorithmic}[1]
		\REQUIRE $\epsilon_{\mathrm{SCA}}$,$\epsilon_{\mathrm{FPE}}$, $i_{\mathrm{max}}$, $N_{\mathrm{t}}$, $N_{\mathrm{r}}$, $N_c$, $L_\mathrm{p}$, $L_\mathrm{d}$, $P_\mathrm{p}$, $P_\mathrm{d}$, $\sigma_\mathrm{s}^2$, $\sigma_\mathrm{c}^2$, $\mathbf{R}$, and $\mathbf{H}_c$.
		\ENSURE A locally optimal solution $\mathbf{M}$ to problem (\ref{ISAC_prob_M}).
		\STATE Initialize $i \gets 0$, $\Delta_{J} \gets \infty$, and $\mathbf{M}_\mathrm{}^{(0)} \gets \frac{P_\mathrm{d}}{N_{\mathrm{t}}} \mathbf{I}_{N_{\mathrm{t}}}$.
		
		\WHILE{$\Delta_{J} \ge \epsilon_{\mathrm{SCA}}$ and $i \le i_{\mathrm{max}}$}
		
		\STATE Initialize $t\gets0$, $\Delta_e \gets \infty$, and calculate $e_0$ via (\ref{e0});
			\WHILE {$\Delta_e \ge \epsilon_{\mathrm{FPE}}$}
			\STATE Calculate $e_{t+1}$ via (\ref{iter_e_eq});
			\STATE Update $\Delta_e \gets \lvert e_{t+1} - e_{t} \rvert$;
			\STATE Update $t \gets t + 1$;
				\ENDWHILE
		\STATE Calculate the current objective function value $J_{\mathrm{ae}}(\mathbf{M}^{(i)})$ via (\ref{eq:Jae});
		\STATE Solve subproblem (\ref{SCA_subprob}) to obtain $\mathbf{M}^{(i+1)}$ by CVX;
    	\STATE Calculate the updated objective function value $J_{\mathrm{ae}}(\mathbf{M}^{(i+1)})$ via (\ref{eq:Jae});	
		\STATE Update $\Delta_{J} \gets  \lvert J_{\mathrm{ae}}(\mathbf{M}^{(i+1)}) - J_{\mathrm{ae}}(\mathbf{M}^{(i)}) \rvert $;
		
		\STATE Update $i \gets  i + 1$;
		
    		\ENDWHILE
	\end{algorithmic}
\end{algorithm}

However, it can be observed that in \textbf{Algorithm \ref{alg:SAC_ISAC}}, for each convex approximation subproblem  in the $i$-th iteration, we need to iteratively solve the fixed-point equation (\ref{e_expression}) to compute the degrading factor $\alpha$, which entails high computational complexity.
In the next section, we will derive a more compact expression for the degrading factor $\alpha$ under the high SNR condition to avoid solving the fixed-point equation, thereby further simplifying the design of the precoding matrix.

\section{ISAC Precoding Design in High SNR Regimes}\label{sec:ISAC_high_SNR}

By taking a closer look at the semi-closed-form asymptotic expression $J_{\mathrm{ae}}$ derived in (\ref{eq:Jae}), we may observe that parameter $\alpha$, which quantifies the sensing performance degradation caused by random data signals, plays a significant role in performance evaluation and precoding design.
Unfortunately, $\alpha$ is implicitly determined by the fixed-point equation (\ref{e_expression}), limiting our ability to derive structural insights into ELMMSE. 
Nevertheless, in this section we demonstrate that in high SNR regions, $\alpha$ can be explicitly expressed in terms of $N_\mathrm{t}$ and $L_\mathrm{d}$, yielding a closed-form expression. 
This result considerably simplifies the derivation of the globally optimal solution to the ISAC precoding optimization problem in the high SNR regions.

\subsection{Asymptotic Expression of ELMMSE at High SNRs}
We start by deriving the closed-form asymptotic expression of $e$ in the high SNR regions, which is detailed in the following proposition.

\begin{prop}\label{lemma:e_HighSNR}
	In the high SNR regions, i.e., $P_{\mathrm{d}}L_{\mathrm{d}}/\sigma_\mathrm{s}^2 \rightarrow \infty$, the asymptotic expression $\bar{e}$, which is defined as 
	\begin{equation}
	\begin{aligned}
  		\bar{e} = \lim_{P_{\mathrm{d}}L_{\mathrm{d}}/\sigma_\mathrm{s}^2 \rightarrow \infty} e,
  	\end{aligned}
	\end{equation}
is given by
	\begin{equation}
	\begin{aligned}
  		\bar{e} = \frac{L_{\mathrm{d}}}{L_{\mathrm{d}}- N_{\mathrm{t}}}.
  	\end{aligned}
	\end{equation}
\end{prop}

\begin{IEEEproof}
	Denote 
	\begin{equation}
	\begin{aligned}
  		\frac{1}{N_{\mathrm{r}} \sigma_\mathrm{s}^2}\mathbf{W}_\mathrm{} \mathbf{W}_\mathrm{}^H = \gamma {\mathbf{\Phi}},
  	\end{aligned}
\end{equation}
where $\gamma = \mathrm{Tr}(\mathbf{W}_\mathrm{} \mathbf{W}_\mathrm{}^H) / N_{\mathrm{r}} \sigma_\mathrm{s}^2$. The fixed-point equation defined in (\ref{e_expression}) can be reformulated as
\begin{equation}\label{highSNR_e}
    \begin{split}
      e =  \frac{1}{N_{\mathrm{t}} } \mathrm{Tr}  \biggl[ \frac{1}{\gamma} {\mathbf{\Phi}}^{-1} \left( \mathbf{R}^{-1} +  \frac{ P_\mathrm{p} } {N_\mathrm{t}N_\mathrm{r} \sigma_{\mathrm{s}}^2} \mathbf{I}_{N_{\mathrm{t}}} \right) 
         +  \frac{L_\mathrm{d}}{L_\mathrm{d} + N_\mathrm{t} e } \mathbf{I}_{N_{\mathrm{t}}}  \biggr]^{-1}.
    \end{split}
\end{equation}
Under the high SNR condition, we have $\gamma \rightarrow \infty$. Note that
\begin{equation}
	\begin{aligned}
  		\bar{e} &= \lim_{\substack{\gamma \rightarrow \infty }} e,
  	\end{aligned}
\end{equation}
and therefore, by taking the limit on both sides of (\ref{highSNR_e}), we obtain
\begin{equation}
	\begin{aligned}
  		\bar{e} = \frac{L_\mathrm{d} + N_\mathrm{t} \bar{e} }{L_\mathrm{d}},
  	\end{aligned}
\end{equation}
which immediately yields 
\begin{equation}
	\begin{aligned}
  		\bar{e} = \frac{L_{\mathrm{d}}}{L_{\mathrm{d}}- N_{\mathrm{t}}}.
  	\end{aligned}
	\end{equation}
\end{IEEEproof}

In high SNR regions, by leveraging Proposition \ref{lemma:e_HighSNR} directly, the factor $\alpha$ defined in (\ref{alpha_def}) can be further expressed as
\begin{equation}\label{alpha_highSNR}
	\begin{aligned}
  		\alpha = 1-\frac{N_{\mathrm{t}}}{L_{\mathrm{d}}}.
  	\end{aligned}
\end{equation}

\emph{Remark 4:} In particular, (\ref{alpha_highSNR}) suggests that factor $\alpha$, which quantifies the sensing performance degradation caused by random signals, is specified as the constant $1-{N_{\mathrm{t}}} / {L_{\mathrm{d}}}$ under the high SNR condition. 
It should be noted that $\alpha$ is positive since $L_{\mathrm{d}} > N_{\mathrm{t}}$.
This result further demonstrates that, when the number of transmit antennas is fixed, high SNR cannot fully compensate for the degradation in sensing performance induced by signal randomness. Instead, only sufficiently long data symbols can achieve negligible performance loss.

Under the high SNR condition, $J_{\mathrm{ae}}$ exhibits a simplified closed-form expression for ELMMSE without the need to solve the fixed-point equation, which is given by

\begin{equation}\label{Jae_high_SNR}
	\begin{aligned}
  		\tilde{J}_{\mathrm{ae}} =\mathrm{Tr} \biggl[ \mathbf{R}^{-1} + \frac{1}{N_{\mathrm{r}} \sigma_\mathrm{s}^2} \biggl(\frac{P_\mathrm{p}}{N_{\mathrm{t}}} \mathbf{I}_{N_{\mathrm{t}}}   
  					+ \left( 1-\frac{N_{\mathrm{t}}}{L_{\mathrm{d}}} \right) \mathbf{W}_\mathrm{} \mathbf{W}_\mathrm{}^H \biggr) \biggr] ^{-1}.
  	\end{aligned}
\end{equation}

{\color{black}{\emph{Remark 5:} In contrast to the semi-closed-form asymptotic expression ${J}_{\mathrm{ae}}$ in (\ref{eq:Jae}), the closed-form asymptotic expression $\tilde{J}_{\mathrm{ae}}$ given in (\ref{Jae_high_SNR}) offers twofold advantages. 
First, $\tilde{J}_{\mathrm{ae}}$ theoretically demonstrates that in high SNR regions, the asymptotic expression of ELMMSE shares an identical structure with its lower bound (\ref{eq:lb_ELMMSE}). 
The only distinction lies in the modification of the scaling factor for matrix $\mathbf{W}\mathbf{W}^H$, where the original coefficient 1 is replaced by a constant $\alpha$ as specified in (\ref{alpha_highSNR}). 
This constant clearly captures how the randomness of data signal degrades sensing performance.
Therefore, compared with the lower bound (\ref{eq:lb_ELMMSE}), which is obtained by directly treating the random signal as a deterministic signal, the closed-form asymptotic expression (\ref{Jae_high_SNR}) significantly improves the approximation accuracy of the ELMMSE.
Second, $\tilde{J}_{\mathrm{ae}}$'s convexity with respect to $\mathbf{W}\mathbf{W}^H$ guarantees the convergence to the globally optimal solution to the ISAC precoding problem in high SNR regions, and provides significant computational advantages. Particularly, it avoids computationally demanding procedures, including fixed-point equation solving and convex approximation steps, in \textbf{Algorithm \ref{alg1}}. 
}}

\subsection{Optimal Precoding Design at High SNRs}
Next, we introduce the ISAC precoding scheme based on the derived asymptotic expression (\ref{Jae_high_SNR}) of ELMMSE in the high SNR regions. Similar to Section \ref{subsec:SCA}, by denoting $\mathbf{M}=\mathbf{WW}^H$, the ISAC precoding optimization problem under the high SNR condition can be formulated as
\begin{equation}\label{ISAC_prob_M_highSNR}
	\begin{aligned}
     \min_{\mathbf{M}_\mathrm{}} \ & \tilde{J}_{\mathrm{ae}}(\mathbf{M})   \\
     \mathrm{s.t.}  \ &   \frac{L_{\mathrm{d}}}{L} \log \det \left(\mathbf{I}_{N_\mathrm{c}}   + \frac{\mathbf{H}_\mathrm{c}  {\mathbf{M}} \mathbf{H}_\mathrm{c}^H}{\sigma_\mathrm{c}^2}\right) \ge R_0,\\
     				\ &    \mathrm{Tr}(\mathbf{M}) \le P_\mathrm{d}, \\
     				\ & \mathbf{M} \succeq \mathbf{0},
  	\end{aligned} 
\end{equation}
where
\begin{equation}\label{highSNR_ISAC_prob_Jae}
	\begin{aligned}
  	 \tilde{J}_{\mathrm{ae}}(\mathbf{M})	= \mathrm{Tr} \biggl[ \mathbf{R}^{-1} + \frac{1}{N_{\mathrm{r}} \sigma_\mathrm{s}^2} \biggl(\frac{P_\mathrm{p}}{N_{\mathrm{t}}} \mathbf{I}_{N_{\mathrm{t}}}
  					+ \left(1-\frac{N_{\mathrm{t}}}{L_{\mathrm{d}}}\right) \mathbf{M} \biggr) \biggr] ^{-1}.
  	\end{aligned}
\end{equation}

It can be observed that the derived closed-form expression $\tilde{J}_{\mathrm{ae}}(\mathbf{M})$ eliminates the calculation of the parameter $e$ which was determined by the fixed-point equation (\ref{e_expression}) in the low SNR cases, thereby converting the originally non-convex ISAC problem (\ref{ISAC_prob_M}) into a convex problem (\ref{ISAC_prob_M_highSNR}), which can be effectively solved using numerical tools such as CVX~\cite{gb08}.
Once the globally optimal solution $\mathbf{M}$ of problem (\ref{ISAC_prob_M_highSNR}) is obtained, denote its eigenvalue decomposition as $\mathbf{M}=\mathbf{Q}\mathbf{\Sigma}\mathbf{Q}^H$, we can formulate the corresponding optimal precoding matrix $\mathbf{W}$ by (\ref{eq:decom_M}).

\subsection{Complexity Analysis}

\setcellgapes{2pt}
\begin{table}[t]
	\centering
	\caption{Complexity Comparisons of Different Algorithms.}
	\label{tab:complexity_comparison}
	\makegapedcells
	\begin{threeparttable}
		\begin{tabular}{c|c|c}
			\hline
			\hline
			 \makecell{ \\}  &\makecell{\bf{SCA-based} } &\makecell{\bf{High SNR Approx.}}  \\ 
			 \hline
			 
			\bf{\bf{Complexity} }	                  &${\cal{O}}(\mathcal{I}_{\mathrm{SCA}}(N_{\mathrm{t}}^{3.25} + \mathcal{I}_{\mathrm{FPE}} N_{\mathrm{t}}^{3}))$	& ${\cal{O}}(N_{\mathrm{t}}^{3.25})$          \\

			\hline                           
			
			\hline
		\end{tabular}
			\end{threeparttable}
\end{table}

Table \ref{tab:complexity_comparison} summarizes the complexity of the proposed SCA-based algorithm and the method based on the high SNR approximation.
For the proposed SCA-based algorithm, the computational complexity is attributed to solving the fixed-point equation (\ref{e_expression}), derivative (\ref{eq:grad_Jae}), and convex approximation problem (\ref{SCA_subprob}).
Denote the iteration numbers of the SCA algorithm and the fixed-point equation-solving algorithm to reach convergence is $\mathcal{I}_{\mathrm{SCA}}$ and $\mathcal{I}_{\mathrm{FPE}}$, respectively. 
The computational complexity of solving convex problems using interior point method in CVX is ${\cal{O}}(N_{\mathrm{t}}^{3.25})$~\cite{jiang2020faster}, while the complexity of solving fixed-point equations and computing gradients is ${\cal{O}}(\mathcal{I}_{\mathrm{FPE}}N_\mathrm{t}^3)$ and ${\cal{O}}(N_\mathrm{t}^3)$, respectively. 
Consequently, the overall computational complexity of the SCA-based algorithm amounts to ${\cal{O}}(\mathcal{I}_{\mathrm{SCA}}(N_{\mathrm{t}}^{3.25} + \mathcal{I}_{\mathrm{FPE}} N_{\mathrm{t}}^{3}))$. 
 For the method at high SNRs, since it requires solving a convex problem only once, its computational complexity is simply ${\cal{O}}(N_{\mathrm{t}}^{3.25})$.
  This again confirms the superiorities of the derived closed-form asymptotic expression (\ref{Jae_high_SNR}) for ELMMSE in the high SNR regimes.

\section{Simulation Results}\label{sec:sim_res}
In this section, we validate the accuracy of the derived asymptotic expressions of ELMMSE and the ISAC performance of proposed optimization methods via numerical simulations. 
We consider a MIMO system with $N_{\mathrm{t}}=N_{\mathrm{r}}=32$ antennas while the communication user is equipped with $N_\mathrm{c}=4$ antennas.
The signal frame length $L=320$.
For the proposed SCA-based algorithm, we set the tolerance $\epsilon_{\mathrm{SCA}}=10^{-3}$ and $\epsilon_{\mathrm{FPE}}=10^{-8}$, respectively. 
The maximum number of iterations was set to $i_{\mathrm{max}}=200$. 
Unless otherwise specified, SNR specifically refers to the signal-to-noise ratio at the transmitter end, and the noise power $\sigma_{\mathrm{s}}^2$ and $\sigma_{\mathrm{c}}^2$ are both set to 0 dBm. 

\subsection{Validation of the Asymptotic Expressions of ELMMSE}

\begin{figure}
	\centering
	\hspace{0cm}\includegraphics[width=0.4\textwidth]{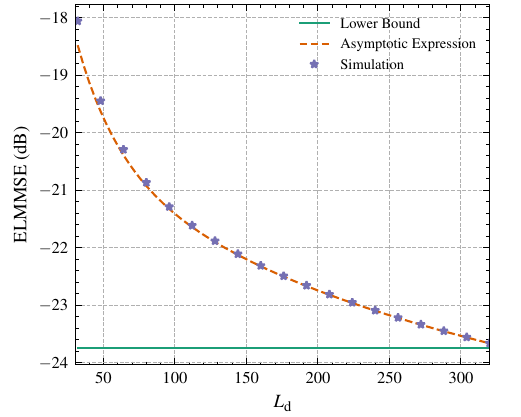}
	\caption{ELMMSE, its lower bound $J_{\mathrm{lb}}$, and approximate expression $J_{\mathrm{ae}}$, versus the length of data symbols $L_\mathrm{d}$.}
	\vspace{1mm}
	\label{fig:convg_Jae}
\end{figure}

\begin{figure}[t]
	\centering
	\vspace{-1mm}\includegraphics[width=0.4\textwidth]{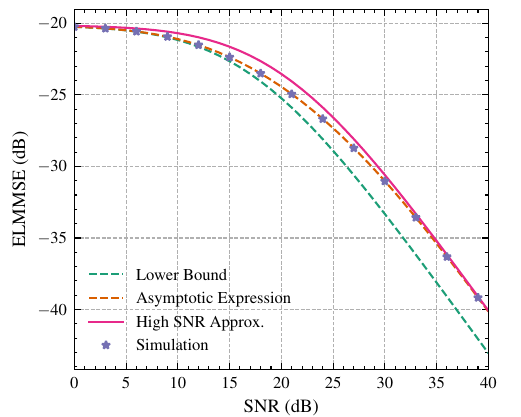}
	\vspace{1mm}
	\caption{ELMMSE, its lower bound $J_{\mathrm{lb}}$, and two approximate expressions $J_{\mathrm{ae}}$ and $\tilde{J}_{\mathrm{ae}}$, versus the transmit SNR.}
	\label{fig:convg_Jae_highSNR}
	\vspace{1mm}
\end{figure}

Fig. \ref{fig:convg_Jae} demonstrates the validity of the asymptotic expressions derived in Section \ref{sec:AE}. 
The legend \enquote{Lower Bound} and \enquote{Asymptotic Expression} correspond to the lower bound (\ref{eq:lb_ELMMSE}) and asymptotic expression (\ref{eq:Jae}) of ELMMSE, respectively, while the legend \enquote{{Simulation}} represents the numerical results of ELMMSE, as defined in (\ref{ELMMSE}), obtained through Monte-Carlo simulations over 5,000 realizations. 
It can be observed that, although in theory, ${J}_{\mathrm{ae}}$ can approximate ELMMSE with arbitrarily low error only when $L_\mathrm{d} \rightarrow \infty$, the simulation results demonstrate that even for a small value of $L_\mathrm{d}$, such as $L_\mathrm{d}=64$, ${J}_{\mathrm{ae}}$ can still approximate ELMMSE with high accuracy.
Furthermore, as $L_{\mathrm{d}}$ increases, ELMMSE gradually decreases and approaches the lower bound ${J}_{\mathrm{lb}}$, as mentioned in Remark 1.
This simulation result indicates that when the random signal length is insufficient, using the lower bound ${J}_{\mathrm{lb}}$ to approximate the ELMMSE will introduce significant errors, further demonstrating the necessity of the derived asymptotic expression ${J}_{\mathrm{ae}}$.

Fig. \ref{fig:convg_Jae_highSNR} validates the accuracy of ${J}_{\mathrm{ae}}$ and its high SNR counterpart $\tilde{J}_{\mathrm{ae}}$ across varying SNR levels. 
The legend \enquote{High-SNR Approx.} corresponds to $\tilde{J}_{\mathrm{ae}}$ derived in (\ref{Jae_high_SNR}). 
It can be observed that when the SNR exceeds 30 dB, the approximation gap introduced by $\tilde{J}_{\mathrm{ae}}$ becomes essentially negligible. 
This indicates that in high SNR regions, the simplified expression $\tilde{J}_{\mathrm{ae}}$ serves as an outstanding alternative to the more complex expression ${J}_{\mathrm{ae}}$, which facilitates efficient design of the optimal precoding matrix.

  \subsection{Precoding Design}

In this subsection, we first evaluate the communication and sensing performance of the proposed ISAC precoding design, which employs both pilot and data signals for sensing tasks, under different system settings. 
The baseline scheme, which utilizes only pilot resources for sensing, is implemented for comparison.

  \begin{figure}[t]
	\centering
	\includegraphics[width=0.4\textwidth]{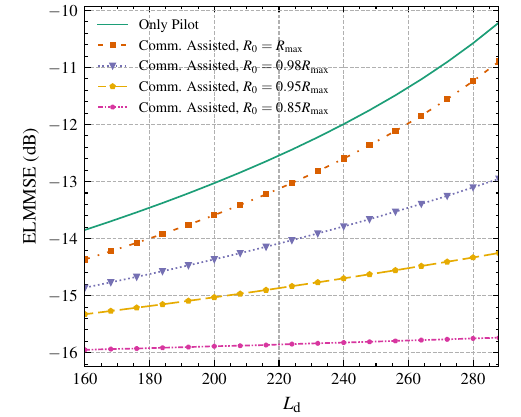}
	\hspace{5mm}
	\caption{ELMMSE comparison among different settings versus $L_\mathrm{d}$ when SNR$= 30$ dB and $L=320$.}
	\label{fig:ELMMSE_5dBm}
	\vspace{3mm}
\end{figure}

 \begin{figure}[t]
	\centering
	\vspace{0cm}
	\includegraphics[width=0.4\textwidth]{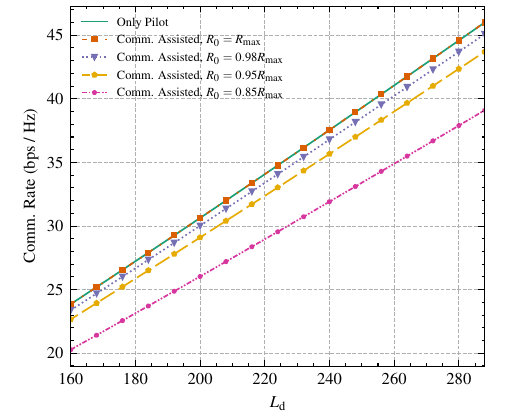}
	\vspace{0mm}
	\caption{Communication rate comparison among different settings versus $L_\mathrm{d}$ when SNR$=30$ dB, $L=320$.}
	\label{fig:Rate_5dBm}
	\vspace{1mm}
\end{figure}

 \begin{figure}[t]
	\centering
	\hspace{-4mm}
	\includegraphics[width=0.4\textwidth]{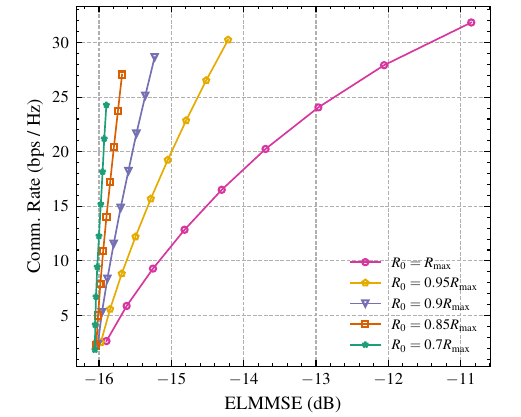}
	\vspace{0mm}
	\caption{Sensing and communication performance tradeoff under different rate thresholds when SNR$=30$ dB, $L=320$. For each curve, the variation in sensing and communication performance is determined by the resource proportion of data symbols. Specifically, the left and right points correspond to the case of $L_{\mathrm{d}}/L = 0.1$ and $L_{\mathrm{d}}/L = 0.9$, respectively.}
	\label{fig:Tradeoff_charecter}
	\vspace{1mm}
\end{figure}

In Fig. \ref{fig:ELMMSE_5dBm} and Fig. \ref{fig:Rate_5dBm}, the ELMMSE and communication performance are shown as functions of the data symbol length when the frame length $L=320$ and SNR $= 30$ dB.
The legend \enquote{Only Pilot} and \enquote{Comm. Assisted} denote the baseline scheme and proposed SCA-based precoding scheme, respectively.
To illustrate the tradeoff between sensing and communication performance, the rate thresholds $R_0$ are set as $R_{\mathrm{max}}$, 0.98$R_{\mathrm{max}}$, 0.95$R_{\mathrm{max}}$, and 0.8$R_{\mathrm{max}}$, respectively, where $R_{\mathrm{max}}$ is the maximum achievable communication rate implemented by the water-filling algorithm.
 
It can be observed in Fig. \ref{fig:ELMMSE_5dBm} that by adjusting the rate threshold $R_0$, varying degrees of tradeoff between sensing and communication performance can be achieved. 
Specifically, when SNR is 30 dB, if $R_0=R_\mathrm{max}$ (indicating that we target at the maximum communication performance), reusing the time-frequency resources of data symbols to assist sensing can reduce the ELMMSE by approximately 0.7 dB. 
When the required minimum communication rate is set to 0.98$R_\mathrm{max}$, a sensing performance gain of approximately 2.8 dB can be achieved. 
Furthermore, by relaxing the communication rate requirement to 0.95$R_\mathrm{max}$ and 0.85$R_\mathrm{max}$, maximum gains of 4 dB and 5.6 dB in sensing performance can be obtained, respectively. 
 These simulation results also clearly demonstrate the advantages of repurposing random data symbols for sensing in ISAC systems. 
 
   \begin{figure}[t]
	\centering
	\vspace{0mm}
	\hspace{0mm}\includegraphics[width=0.4\textwidth]{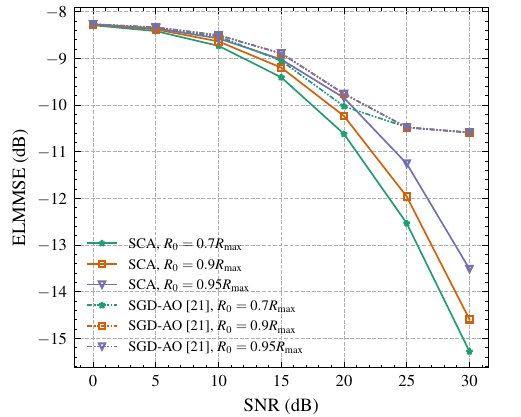}
	\vspace{0mm} 
	\caption{ELMMSE comparison among different precoding schemes versus transmit SNR when $L_{\mathrm{d}}=$ 96.}
	\label{fig:comm_my_lu}
	 \vspace{-1mm}
\end{figure} 
 
 {\color{black}{Generally, for deterministic pilot and random data signals of equal length and power, deterministic pilot signals exhibit a superior sensing performance. 
 Consequently, increasing the proportion of random data signals within a signal frame will lead to a degradation in the overall sensing performance of the frame, as shown in Fig. \ref{fig:ELMMSE_5dBm}.}}
 However, when the communication performance requirement is less stringent, more degrees of freedom in precoding design can be exploited for the sensing performance enhancement.
 Notably, as shown in  Fig. \ref{fig:Rate_5dBm}, when the communication rate threshold is constrained to be no less than 85\% of the maximum achievable rate, increasing the data payload resource allocation from 50\% to 90\% (corresponding to an increase in $L_{\mathrm{d}}$ from 160 to 288) enjoys an improvement in the achievable communication rate by 95\% (about 19 bps / Hz), only incurring 0.2 dB loss in ELMMSE.
 {\color{black}{Fig. \ref{fig:Tradeoff_charecter} illustrates the performance tradeoff of the proposed ISAC transmission, where each curve represents a distinct tradeoff characteristic.
 As previously discussed, appropriately relaxing the rate threshold $R_0$ can effectively mitigate the performance tradeoff caused by the varying resource proportion of the data symbols. This means that optimizing communication performance will no longer significantly degrade sensing performance, which is a highly desirable characteristic for ISAC systems.
 For example, when $R_0$ is relaxed from $R_{\mathrm{max}}$ to $0.95R_{\mathrm{max}}$, a maximum sensing performance gain of approximately 3 dB can be achieved, while the communication rate only decreases by 1.7 bps/Hz (about 5\%).}} 
{\color{black}{These simulation results demonstrate that exploiting the random data symbols both for sensing and communication, our proposed precoding method can significantly reduce the sensing error while maintaining negligible communication performance loss. This approach thereby presents a viable technical solution for communication-centric ISAC systems.}}

 Fig. \ref{fig:comm_my_lu} compares the sensing performance of our proposed precoding scheme with penalty-based stochastic gradient projection (SGP) alternating optimization (AO) algorithm proposed in \cite{10596930}. 
 The legend \enquote{SCA} and \enquote{SGD-AO} denote the proposed SCA-based algorithm and benchmark technique, respectively.
 To align the system setting of \cite{10596930} for fair comparison, we disregard pilot signals and only consider the sensing performance achieved through random data signals.
 Remarkably, our proposed precoding scheme outperforms the SGP-AO algorithm proposed in \cite{10596930} over all SNR regions.
 {\color{black}{As the SNR increases, the sensing error of the proposed scheme gradually decreases, whereas the performance of the SGD-AO algorithm exhibits a pronounced saturation effect in the high SNR regions. 
 Specifically, taking the case where the communication rate threshold $R_0 = 0.95 R_\mathrm{max}$ as an example, when the SNR increases from 20 dB to 30 dB, the ELMMSE of the proposed SCA-based algorithm decreases from $-$9.9 dB to $-$13.5 dB, while the ELMMSE of the SGD algorithm only decreases from $-$9.8 dB to $-$10.6 dB.}}
 This phenomenon occurs primarily because the performance of penalty-based algorithm in \cite{10596930} is critically constrained by the selection of penalty coefficients. 
 Specifically, an excessively small penalty coefficient may result in constraint violations, while an overly large one will trap the solution in a highly suboptimal solution \cite{penaltyopt_Ksenija_2024}.
 Therefore, determining optimal coefficients constitutes a non-trivial challenge.
 However, in our proposed algorithm, although the asymptotic expression $J_{\mathrm{ae}}$ derived from RMT remains a non-convex function of the precoding matrix $\mathbf{W}$, we can construct a convex function to approximate $J_{\mathrm{ae}}$ using its gradient.
This approach yields a locally optimal solution by leveraging the SCA technique. 
Notably, the communication rate constraint is strictly satisfied in each SCA iteration, and as iterations progress, the solution converges to hit the communication rate target.

 \subsection{Precoding Design in High SNR Regions}
  
     \begin{figure}[t]
	\centering
	\hspace{1mm}\includegraphics[width=0.4\textwidth]{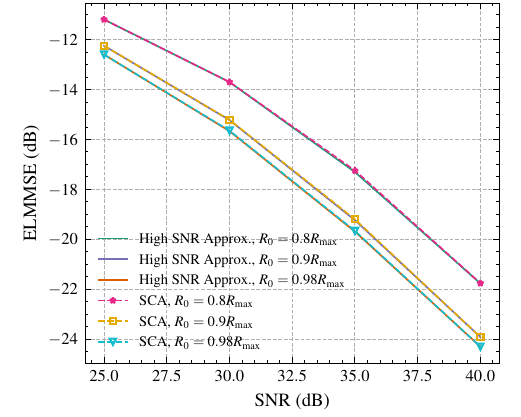}
	\vspace{0mm}
	\caption{ELMMSE comparison between the SCA-based and high-SNR approximation-based method versus transmit SNR, when $L_\mathrm{p}=$ 32 and $L_\mathrm{d}=$ 96.}
	\label{fig:highSNR_comp}
	\vspace{0mm}  
\end{figure}

In this section, the ISAC performance of the proposed precoding design scheme is investigated under the high SNR scenarios.
The derived closed-form asymptotic expression of ELMMSE transfers the originally non-convex ISAC precoding optimization problem into a convex one, enabling efficient computation of its globally optimal solution using the numerical optimization tools CVX.
Fig. \ref{fig:highSNR_comp} compares the sensing performance of different precoding schemes in high SNR regions, where the legend \enquote{High SNR Approx.} denotes the precoding design leveraging the high SNR approximation expression $\tilde{J}_{\mathrm{ae}}$ derived in (\ref{Jae_high_SNR}).
It can be observed that the sensing performance of both precoding schemes remains almost identical under varying SNR conditions or different communication rate constraint thresholds. 
{\color{black}{This result indicates that in the high SNR regions, the derived closed-form expression (\ref{Jae_high_SNR}) can accurately evaluate ELMMSE and, in turn, implies that the proposed SCA-based algorithm can achieve a near-optimal solution.}}


\subsection{Comparison of Runtime}

Table \ref{tab:runtime} compares the average runtime of the SCA-based algorithm, the method based on the high SNR approximation, and the SGP-AO algorithm~\cite{10596930}.
It can be observed that the proposed SCA-based algorithm reduces the average running time by approximately 27\% compared to the penalty-based SGD-AO algorithm in \cite{10596930}.
The method based on the high SNR approximation achieves nearly identical sensing performance while reducing the runtime to approximately only 1\% of that of the SCA-based algorithm.
This is mainly because the SGD-AO algorithm proposed in \cite{10596930} calculates {\color{black}{a suboptimal}} precoding matrix for different realizations of the random data signals in each iteration. 
It often requires multiple iterations to converge, which is demonstrated in the iterative process as the objective function does not strictly decrease with the iterations.

In contrast, our proposed SCA-based algorithm, leveraging an asymptotic expression derived from RMT, guarantees a strict monotonic decrease in the objective function throughout the optimization process, and therefore achieves superior computational efficiency relative to the SGP-AO approach.
However, the SCA-based algorithm still needs to complete high-complexity tasks such as solving the fixed-point equation and calculating the complex gradient in each iteration. In contrast, the method based on the high SNR approximation only needs to solve a convex problem once, which greatly reduces the computational complexity.\footnote{All algorithms were implemented in MATLAB R2024b and executed on a computing platform equipped with an Intel i7-14700K @ 3.4 GHz CPU and 32 GB RAM.}

\begin{table}[t]
	\centering
	\caption{Runtime Comparisons of Different Algorithms.}
	\label{tab:runtime}
	\begin{threeparttable}
		\begin{tabular}{c|c|c|c}
			\hline
			\hline
			 \makecell{ \\}  &\makecell{\bf{SGD-AO \cite{10596930}}} &\makecell{\bf{SCA-based} } &\makecell{\bf{High SNR Approx.}}  \\ \hline
			\bf{\bf{Runtime (s)} }	           &  $326$       &$210$	& $2.4$          \\ \hline                           
			
			\hline
		\end{tabular}
			\end{threeparttable}

\end{table}

\section{Conclusion}\label{sec:conclusion}

This paper investigated the ISAC performance of a multiple antenna system when both deterministic pilot and random data symbols are employed for sensing tasks.
We started by analyzing the properties of ELMMSE, which evaluates the average sensing error of the ISAC system utilizing random data payload symbols in sensing tasks.
{\color{black}{By leveraging RMT, we derived two (semi-)closed-form asymptotic expressions of ELMMSE, which serve as the theoretical foundation for the development of the proposed efficient precoding schemes.}}
Finally, simulation results demonstrated that reusing the random data signals can significantly enhance ISAC system's sensing performance, while validating the effectiveness of our proposed algorithm for precoding design.
{\color{black}{Furthermore, the simulation results revealed that, under some practical settings, our scheme can significantly enhance the sensing performance of ISAC systems without substantially degrading the communication performance.}}
Based on these findings, we conclude that, by leveraging random data signals for sensing tasks, 6G ISAC systems can significantly enhance environmental sensing capabilities without substantially compromising communication service quality for users. This dual functionality demonstrates its potential for applications such as autonomous driving and the low-altitude economy.

\appendices
\section{Proof of Proposition \ref{prop:lowerBound_ELMMSE}}
\label{sec:appendix1}
According to Jensen's inequality, we have
\begin{equation}
	\begin{aligned}
		& J_{\mathrm{ELMMSE}}  =   \mathbb{E}_{{\mathbf{S}}_\mathrm{d}} \biggl\{ \mathrm{Tr} \biggl[ \mathbf{R}^{-1} +  \frac{1}{N_{\mathrm{r}} \sigma_\mathrm{s}^2} \biggl(  \frac{P_\mathrm{p}}{N_{\mathrm{t}}} \mathbf{I}_{N_{\mathrm{t}}} \\
       &  \quad \quad \quad \quad \quad \quad \quad \quad \quad \quad +   \mathbf{W}_\mathrm{} \mathbf{S}_\mathrm{d} \mathbf{S}_\mathrm{d}^H \mathbf{W}_\mathrm{}^H  \biggr) \biggr]^{-1} \biggr\} \ge \\
	&	\mathrm{Tr} \left( \mathbf{R}^{-1} + \frac{1}{N_{\mathrm{r}} \sigma_\mathrm{s}^2} \biggl(  \frac{P_\mathrm{p}}{N_{\mathrm{t}}} \mathbf{I}_{N_{\mathrm{t}}}  + \mathbf{W}_\mathrm{p} \mathbb{E} \left[\mathbf{S}_\mathrm{d} \mathbf{S}_\mathrm{d}^H \right] {\mathbf{W}}_\mathrm{p}^H \biggr)  \right) ^{-1}.
	\end{aligned}
\end{equation}
Recall that the entries of $\mathbf{S}_\mathrm{d}$ are i.i.d. complex Gaussian variables with zero mean and variance $1/{L_\mathrm{d}}$, and therefore we have
\begin{equation}
	\mathbb{E} [{\mathbf{S}}_\mathrm{d} {\mathbf{S}}_\mathrm{d}^H] = \mathbf{I}_{N_\mathrm{t}}.
	\end{equation}
Consequently, denote $J_{\mathrm{lb}}$ as the lower bound of $J_{\mathrm{ELMMSE}}$, and we have
\begin{equation}
	\begin{aligned}
  			J_{\mathrm{ELMMSE}} & \ge \mathrm{Tr} \left[ \mathbf{R}^{-1} + \frac{1}{N_{\mathrm{r}} \sigma_\mathrm{s}^2} \left( \frac{P_\mathrm{p}}{N_{\mathrm{t}}} \mathbf{I}_{N_{\mathrm{t}}} + \mathbf{W}_\mathrm{} \mathbf{W}_\mathrm{}^H \right) \right] ^{-1} \\
  			& = J_{\mathrm{lb}},
  	\end{aligned}
\end{equation}
which completes the proof.
\qed

\section{Proof of Proposition \ref{prop:AE_ELMMSE}}
\label{sec:appendix_AE_ELMMSE}

Recall the definition of the matrix $\mathbf{G}$
\begin{equation}
	\mathbf{G} = {\mathbf{A}} + {\mathbf{BSS}}^H {\mathbf{B}}^H,
\end{equation}
where ${\mathbf{A}} \in \mathbb{C}^{N_\mathrm{} \times N_\mathrm{}}$ is a non-negative definite matrix, and ${\mathbf{B}} \in \mathbb{C}^{N_\mathrm{} \times N_\mathrm{}}$ is an arbitrary square matrix. ${\mathbf{S}} \in \mathbb{C}^{N_\mathrm{} \times M_\mathrm{}} $ is a random matrix with zero mean and variance $1/M_\mathrm{}$ i.i.d. entries. 
The prerequisite for Theorem {\ref{theorem:tit} to hold is that $\mathbf{B}$ is a square root of a non-negative definite matrix. We first demonstrate that Theorem  {\ref{theorem:tit} holds for any $\mathbf{B}$. 
Denote the eigenvalue decomposition of $\mathbf{B}_\mathrm{} \mathbf{B}_\mathrm{}^H$ as

\begin{equation}
	\begin{aligned}
  		\mathbf{B}_\mathrm{} \mathbf{B}_\mathrm{}^H = \mathbf{Q}\mathbf{\Lambda}\mathbf{Q}^H,
  	\end{aligned}
\end{equation} 
where $\mathbf{Q}$ is an orthogonal matrix, $\mathbf{\Lambda}$ is a diagonal matrix with all diagonal entries non-negative. Therefore, $\mathbf{B}_\mathrm{}$ can be expressed as
\begin{equation}
	\begin{aligned}
  		\mathbf{B}_\mathrm{} = \mathbf{Q}\mathbf{\Lambda}^{\frac{1}{2}},
  	\end{aligned}
\end{equation}
and ${\mathbf{G}}$ can be recast to
\begin{equation}
	\begin{aligned}
  		{\mathbf{G}} = {\mathbf{A}} +\mathbf{Q} \mathbf{\Lambda}^{\frac{1}{2}} \mathbf{Q}^H(\mathbf{QS})(\mathbf{S}^H\mathbf{Q}^H) \mathbf{Q} \mathbf{\Lambda}^{\frac{1}{2}} \mathbf{Q}^H,
  	\end{aligned}
\end{equation}
note that $\mathbf{QS}$ remains a Gaussian random matrix due to the orthogonality of $\mathbf{Q}$.
Therefore, according to Theorem {\ref{theorem:tit}, we have
\begin{equation}
	\begin{aligned}
  		m_{N_{\mathrm{}}}^{\mathbf{G}}(z) - m^{\mathbf{G}}(z) \xrightarrow[]{N,M\to \infty}   0
  	\end{aligned}
\end{equation}
with $N_{\mathrm{}} / M_{\mathrm{}} \to c\in (0,\infty)$.
According to (\ref{eq:esd}) and (\ref{eq:stieltjes_tsfm}), the Stieltjes transform $m_{N_{\mathrm{}}}^{\mathbf{G}}(z)$ at $z=0$ can be expressed as
\begin{equation}\label{recast_st}
	\begin{aligned}
  		m_{N_{\mathrm{}}}^{\mathbf{G}}(0) = \frac{1}{N_{\mathrm{}}} \mathrm{Tr}\mathbf{G} ^{-1}.
  	\end{aligned}
\end{equation}
According to (\ref{eq:tit_mz}), $m^{\mathbf{G}}(0)$ is given by
\begin{equation}\label{mG0}
  		m^{\mathbf{G}}(0)= \frac{1}{N_{\mathrm{}}} \mathrm{Tr} \left(  \mathbf{A} +  \frac{\mathbf{Q}\mathbf{\Lambda}\mathbf{Q}^H} {1+c  e}   \right)^{-1},
\end{equation}
where $e$ is the solution to the following fixed-point equation
\begin{equation}\label{fpe_e}
	\begin{aligned}
  		e = \frac{1}{N_{\mathrm{}}} \mathrm{Tr} \left[ \mathbf{Q}\mathbf{\Lambda}\mathbf{Q}^H \left(  \mathbf{A} +  \frac{\mathbf{Q}\mathbf{\Lambda}\mathbf{Q}^H} {1+c  e}  \right)^{-1} \right] .
  	\end{aligned}
\end{equation}
Therefore we have 
\begin{equation}
	\begin{aligned}
  		\mathrm{Tr}\mathbf{G} ^{-1} - N m^{\mathbf{G}}(0) \xrightarrow[]{N,M \to \infty}  0
  	\end{aligned}
\end{equation}
with $N_{\mathrm{}} / M_{\mathrm{}} \to c\in (0,\infty)$. Since $\mathrm{Tr}\mathbf{G} ^{-1}$ is uniformly integrable, we have
\begin{equation}\label{aprox_TrG_mG}
	\begin{aligned}
  		\mathbb{E}_{{\mathbf{S}}} \left[ \mathrm{Tr} \mathbf{G}^{-1} \right] - \mathbb{E}_{{\mathbf{S}}} \left[ N m^{\mathbf{G}}(0) \right]  \xrightarrow[]{N,M \to \infty}  0.
  	\end{aligned}
\end{equation}
Recall that $m^{\mathbf{G}}(z)$ is a deterministic function, (\ref{aprox_TrG_mG}) can be rewritten as 
\begin{equation}\label{aprox2_TrG_mG}
	\begin{aligned}
  		\mathbb{E}_{{\mathbf{S}}} \left[ \mathrm{Tr} \mathbf{G}^{-1} \right] -  N m^{\mathbf{G}}(0)  \xrightarrow[]{N,M \to \infty}  0.
  	\end{aligned}
\end{equation}
To derive the asymptotic expression of ELMMSE from (\ref{aprox2_TrG_mG}), we first adopt the following notations, 
\begin{subequations}\label{matrix_notation_RMT}
	\begin{align}
		c &= N_{\mathrm{t}} / L_{\mathrm{d}} \\
		\mathbf{A} &= \mathbf{R}^{-1} + \frac{1}{N_{\mathrm{r}} \sigma_\mathrm{s}^2} \frac{P_\mathrm{p}}{N_{\mathrm{t}}}  \mathbf{I}_{N_{\mathrm{t}}},  \\
  		\mathbf{B} &= \frac{1}{\sqrt{N_{\mathrm{r}} \sigma_\mathrm{s}^2}}\mathbf{W}_{\mathrm{}}, \\
  		\mathbf{S} &= \mathbf{S}_\mathrm{d}, \\
  		\mathbf{G} & = \mathbf{A} +  \mathbf{B} \mathbf{S}\mathbf{S}^H\mathbf{B}^H.
	\end{align}
\end{subequations} 
Consequently, ELMMSE can be reformulated as
\begin{equation}
	\begin{aligned}
  		J_{\mathrm{ELMMSE}} = \mathbb{E}_{{\mathbf{S}}} \left[ \mathrm{Tr} \mathbf{G}^{-1} \right].
  	\end{aligned}
\end{equation}
According to (\ref{aprox2_TrG_mG}), we have
\begin{equation}
	\begin{aligned}
  		J_{\mathrm{ELMMSE}}  - N_{\mathrm{t}} m^{\mathbf{G}}(0)\xrightarrow[]{N_{\mathrm{t}},L_{\mathrm{d}} \to \infty}  0.
  	\end{aligned}
\end{equation}
Substituting (\ref{matrix_notation_RMT}) into (\ref{mG0}) and (\ref{fpe_e}), we have
\begin{equation}
	\begin{aligned}
  		m^{\mathbf{G}}(0) = \frac{1}{N_{\mathrm{t}}} \mathrm{Tr} \biggl[ \mathbf{R}^{-1} + \frac{1}{N_{\mathrm{r}} \sigma_\mathrm{s}^2} \biggl(\frac{P_\mathrm{p}}{N_{\mathrm{t}}}  \mathbf{I}_{N_{\mathrm{t}}}
  					+ \alpha \mathbf{W}_\mathrm{} \mathbf{W}_\mathrm{}^H \biggr) \biggr] ^{-1},
  	\end{aligned}
\end{equation}
where
\begin{equation}
	\begin{aligned}
  		\alpha = \frac{L_{\mathrm{d}}}{L_{\mathrm{d}} + N_{\mathrm{t}} e},
  	\end{aligned}
\end{equation}
and $e$ is the unique solution to the following fixed-point equation
\begin{equation}
	\begin{aligned}
  		e =  \frac{1}{N_{\mathrm{t}} N_{\mathrm{r}}  \sigma_\mathrm{s}^2}  \mathrm{Tr} \biggl\{ \mathbf{W}_\mathrm{} \mathbf{W}_\mathrm{}^H & \biggl[ \mathbf{R}^{-1} + \frac{1}{N_{\mathrm{r}} \sigma_\mathrm{s}^2} \biggl( \frac{P_\mathrm{p}}{N_{\mathrm{t}}}  \mathbf{I}_{N_{\mathrm{t}}}\\
        &  +   \frac{L_{\mathrm{d}}}{L_{\mathrm{d}} + N_{\mathrm{t}} e} \mathbf{W}_\mathrm{} \mathbf{W}_\mathrm{}^H  \biggr) \biggr]^{-1} \biggl\}.
  	\end{aligned}
\end{equation}
To solve this fixed-point equation iteratively, we employ (\ref{ez_convgence}) and (\ref{iter_solve_e}) to express $e$ in the following explicit form, 
\begin{equation}
	\begin{aligned}
  		e = \lim_{t \to \infty}e_t,
  	\end{aligned}
\end{equation}
where 
\begin{equation}
	\begin{aligned} \label{proof_iter_e_eq}
  		e_t  =  \frac{1}{N_{\mathrm{t}} N_{\mathrm{r}}  \sigma_\mathrm{s}^2}  \mathrm{Tr} \biggl\{ \mathbf{W}_\mathrm{} \mathbf{W}_\mathrm{}^H & \biggl[ \mathbf{R}^{-1} + \frac{1}{N_{\mathrm{r}} \sigma_\mathrm{s}^2} \biggl(  \frac{P_\mathrm{p}}{N_{\mathrm{t}}}  \mathbf{I}_{N_{\mathrm{t}}}\\
        &  +   \frac{L_{\mathrm{d}}}{L_{\mathrm{d}} + N_{\mathrm{t}} e_{t-1}} \mathbf{W}_\mathrm{} \mathbf{W}_\mathrm{}^H  \biggr) \biggr]^{-1} \biggl\}
  	\end{aligned}
\end{equation}
for $t \ge 1$. Insert the initial solution $-\frac{1}{z}|_{z=0}$ given by Theorem \ref{theorem:tit} into (\ref{proof_iter_e_eq}), we have
\begin{equation}
	\begin{aligned}
  		e_0 = \frac{1}{N_{\mathrm{t}} N_{\mathrm{r}}  \sigma_\mathrm{s}^2}  \mathrm{Tr} \biggl[ \mathbf{W}_\mathrm{} \mathbf{W}_\mathrm{}^H & \biggl( \mathbf{R}^{-1} + \frac{1}{N_{\mathrm{r}} \sigma_\mathrm{s}^2}   \frac{P_\mathrm{p}}{N_{\mathrm{t}}}  \mathbf{I}_{N_{\mathrm{t}}} \biggr)^{-1} \biggr],
  	\end{aligned}
\end{equation}
thus completes the proof.
\qed

\section{Proof of Proposition \ref{Derivative_J}}\label{sec:appendix_grad_Jae_proof}

Recall the notation in ($\ref{Jae_grad_notation}$), $e$ can be recast to
\begin{equation}\label{A_C_e}
	\begin{aligned}
  		e = a \mathrm{Tr}  (\mathbf{M} \mathbf{T}). 
  	\end{aligned}
\end{equation}
The differential of $e$ is given by
\begin{equation}\label{A_C_de}
	\begin{aligned}
  		\mathrm{d} e = a \mathrm{Tr} [  (\mathrm{d}\mathbf{M})\mathbf{T} + \mathbf{M}(\mathrm{d}\mathbf{T})],
  	\end{aligned}
\end{equation}
where $\mathrm{d}\mathbf{T}$ can be expressed as
\begin{equation}\label{A_C_dT}
	\begin{aligned}
  		\mathrm{d}\mathbf{T} = - a N_{\mathrm{t}} \mathbf{T} \mathrm{d}( \alpha \mathbf{M}) \mathbf{T}.
  	\end{aligned}
\end{equation}
Note that $\alpha = 1 / (1 + e {N_{\mathrm{t}}}/{L_{\mathrm{d}}} )$, we have
\begin{equation}\label{A_C_daM}
	\begin{aligned}
  		\mathrm{d}( \alpha \mathbf{M}) = -\frac{N_{\mathrm{t}}}{L_{\mathrm{d}}} \alpha^2  \mathbf{M} \mathrm{d}e + \alpha \mathrm{d}\mathbf{M}.
  	\end{aligned}
\end{equation}
Insert (\ref{A_C_dT}) and (\ref{A_C_daM}) into (\ref{A_C_de}), we have
\begin{equation}\label{A_C_de_2}
	\begin{aligned}
  		\mathrm{d} e = \frac{1}{\beta} \mathrm{Tr}(\mathbf{G} \mathrm{d}\mathbf{M}),
  	\end{aligned}
\end{equation}
where
\begin{subequations}
	\begin{align}
		\beta &= \frac{1}{a} - \frac{a \alpha^2  N_{\mathrm{t}}^2}{L_{\mathrm{d}}} \mathrm{Tr}(\mathbf{MTMT}), \\
		\mathbf{G} &= \mathbf{T} - a  \alpha N_{\mathrm{t}} \mathbf{TMT}.
  	\end{align}
\end{subequations}
Insert (\ref{A_C_daM}) and (\ref{A_C_de_2}) into (\ref{A_C_dT}), we obtain
\begin{equation}\label{dT}
	\begin{aligned}
  		\mathrm{d}\mathbf{T} =   \left( \frac{a N_{\mathrm{t}}^2 \alpha^2}{\beta L_{\mathrm{d}}} \mathrm{Tr}(\mathbf{T}^2\mathbf{M}) \mathbf{G}-a N_{\mathrm{t}}\alpha \mathbf{T}^2 \right)  \mathrm{d}\mathbf{M}.
  	\end{aligned}
\end{equation}
Recall that $J_{\mathrm{ae}}(\mathbf{M})=\mathrm{Tr}(\mathbf{T})$, we have
\begin{equation}\label{dJ}
	\begin{aligned}
  		\mathrm{d}J_{\mathrm{ae}}(\mathbf{M})=\mathrm{Tr}(\mathrm{d}\mathbf{T}).
  	\end{aligned}
\end{equation}
According to the definition of Wirtinger derivatives, substituting (\ref{dT}) into (\ref{dJ}) we have
\begin{equation}
	\begin{aligned}
  		\frac{\partial J_{\mathrm{ae}}(\mathbf{M})}{\partial \mathbf{M}_\mathrm{}^*}  = \frac{a N_{\mathrm{t}}^2 \alpha^2}{2 \beta L_{\mathrm{d}}} \mathrm{Tr}(\mathbf{T}^2\mathbf{M}) \mathbf{G} - \frac{a N_{\mathrm{t}}\alpha}{2} \mathbf{T}^2,
  	\end{aligned}
\end{equation}
which completes the proof.

 \qed

\bibliographystyle{IEEEtran}
\bibliography{IEEEabrv,references}
\end{document}